\documentclass[journal]{IEEEtran}

\ifCLASSINFOpdf
\else
   \usepackage[dvips]{graphicx}
\fi
\usepackage{url}

\usepackage{graphicx}

\usepackage{cite}
\usepackage{amsmath,amssymb,amsfonts}
\usepackage{algorithmic}
\usepackage{graphicx}                      
\usepackage{textcomp}
\usepackage{psfrag,caption,subcaption}
\usepackage{xcolor}
\usepackage[normalem]{ulem}
\usepackage[ruled,vlined]{algorithm2e}
\usepackage{epstopdf} 
\usepackage{bbm}

\usepackage{amsthm}
\usepackage{algorithmic}%
\usepackage[utf8]{inputenc}
\usepackage[english]{babel}
\usepackage{todonotes}
\newtheorem{theorem}{Theorem}
\newtheorem{lemma}[theorem]{Lemma}

\usepackage{color,soul}
\usepackage{float}

\usepackage{tikz}
\usetikzlibrary{shapes,arrows}
\usetikzlibrary{decorations.markings,positioning,arrows,calc}
\usepackage{schemabloc}
\usetikzlibrary{circuits}

\tikzstyle{block} = [draw, rectangle, 
minimum height=2em, minimum width=3em]
\tikzstyle{sum} = [draw,  circle, node distance=1cm]
\tikzstyle{input} = [coordinate]
\tikzstyle{output} = [coordinate]
\tikzstyle{pinstyle} = [pin edge={to-,thin,black}]

\def\cast{{
   \mathord{
      \hbox to 0em{
         \ooalign{
	   \smash{\hbox{$\ast$}}\crcr
	   \smash{\hskip-1pt\Large\hbox{$\circ$}} }
	 \hidewidth}
      \phantom{\bigcirc}
} }}

\def\bm#1{\mbox{\boldmath $#1$}}

\newcommand{\mPsi}{\mbox{$\bm \Psi$}}

\newcommand{\rH}{^{ \raisebox{1pt}{$\rm \scriptscriptstyle H$}}}
\newcommand{\rHm}{^{\raisebox{1pt}{-$\rm \scriptscriptstyle H$}}}
\newcommand{\rT}{^{ \raisebox{1.2pt}{$\rm \scriptstyle T$}}}

\newcommand{\bds}{\begin {itemize}}
\newcommand{\eds}{\end {itemize}}
\newcommand{\bdf}{\begin{definition}}
\newcommand{\blm}{\begin{lemma}}
\newcommand{\edf}{\end{definition}}
\newcommand{\elm}{\end{lemma}}
\newcommand{\bthm}{\begin{theorem}}
\newcommand{\ethm}{\end{theorem}}
\newcommand{\bprp}{\begin{prop}}
\newcommand{\eprp}{\end{prop}}
\newcommand{\bcl}{\begin{claim}}
\newcommand{\ecl}{\end{claim}}
\newcommand{\bcr}{\begin{coro}}
\newcommand{\ecr}{\end{coro}}
\newcommand{\bquest}{\begin{question}}
\newcommand{\equest}{\end{question}}

\newcommand{\larrow}{{\larrow}}

\newcommand{\argmin}{\ensuremath{\mathrm{arg}\min}}
\newcommand{\argmax}{\ensuremath{\mathrm{arg}\max}}

\newcommand{\cA}{{\ensuremath{\mathcal{A}}}}

\newcommand{\cC}{{\ensuremath{\mathcal{C}}}}
\newcommand{\cD}{{\ensuremath{\mathcal{D}}}}

\newcommand{\cI}{{\ensuremath{\mathcal{I}}}}
\newcommand{\cJ}{{\ensuremath{\mathcal{J}}}}

\newcommand{\cL}{{\ensuremath{\mathcal{L}}}}

\newcommand{\cN}{{\ensuremath{\mathcal{N}}}}

\newcommand{\cQ}{{\ensuremath{\mathcal{Q}}}}

\def\mbC{{\ensuremath{\mathbb C}}}

\def\mbE{{\ensuremath{\mathbb E}}}

\def\mbR{{\ensuremath{\mathbb R}}}

\newcommand{\va}{{\ensuremath{{\mathbf{a}}}}}

\newcommand{\vb}{{\ensuremath{{\mathbf{b}}}}}
\newcommand{\vc}{{\ensuremath{{\mathbf{c}}}}}

\newcommand{\ve}{{\ensuremath{{\mathbf{e}}}}}

\newcommand{\vh}{{\ensuremath{{\mathbf{h}}}}}

\newcommand{\vn}{{\ensuremath{{\mathbf{n}}}}}

\newcommand{\vp}{{\ensuremath{{\mathbf{p}}}}}
\newcommand{\vq}{{\ensuremath{{\mathbf{q}}}}}

\newcommand{\vs}{{\ensuremath{{\mathbf{s}}}}}

\newcommand{\vw}{{\ensuremath{{\mathbf{w}}}}}

\newcommand{\vx}{{\ensuremath{{\mathbf{x}}}}}

\newcommand{\vy}{{\ensuremath{{\mathbf{y}}}}}
\newcommand{\vz}{{\ensuremath{{\mathbf{z}}}}}

\newcommand{\mA}{{\ensuremath{\mathbf{A}}}}

\newcommand{\mB}{{\ensuremath{\mathbf{B}}}}

\newcommand{\mC}{{\ensuremath{\mathbf{C}}}}

\newcommand{\mD}{{\ensuremath{\mathbf{D}}}}

\newcommand{\mE}{{\ensuremath{\mathbf{E}}}}

\newcommand{\mH}{{\ensuremath{\mathbf{H}}}}

\newcommand{\mI}{{\ensuremath{\mathbf{I}}}}

\newcommand{\mN}{{\ensuremath{\mathbf{N}}}}
\newcommand{\mP}{{\ensuremath{\mathbf{P}}}}
\newcommand{\mQ}{{\ensuremath{\mathbf{Q}}}}

\newcommand{\mR}{{\ensuremath{\mathbf{R}}}}

\newcommand{\mS}{{\ensuremath{\mathbf{S}}}}

\newcommand{\mU}{{\ensuremath{\mathbf{U}}}}

\newcommand{\mV}{{\ensuremath{\mathbf{V}}}}

\newcommand{\mX}{{\ensuremath{\mathbf{X}}}}
\newcommand{\mY}{{\ensuremath{\mathbf{Y}}}}
\newcommand{\mZ}{{\ensuremath{\mathbf{Z}}}}

\usepackage{latexsym}

\def\IC{\mathbb C}
\def\IN{\mathbb N}
\def\IZ{\mathbb Z}
\def\IR{\mathbb R}

\def\shat{^{\mathchoice{}{}%
 {\,\,\smash{\hbox{\lower4pt\hbox{$\widehat{\null}$}}}}%
 {\,\smash{\hbox{\lower3pt\hbox{$\hat{\null}$}}}}}}

\def\bSigma{{
      \ooalign{
      \smash{\hskip.4pt\raise.4pt\hbox{$\Sigma$}}\vphantom{}\crcr
      \smash{\hskip.7pt\raise.6pt\hbox{$\Sigma$}}\vphantom{}\crcr
      \smash{\hbox{$\Sigma$}}\vphantom{$\Sigma$}}
      \vphantom{\hbox{$\Sigma$}}
      }}
\def\bTheta{{
      \ooalign{
      \smash{\hskip.5pt\raise.5pt\hbox{$\Theta$}}\vphantom{}\crcr
      \smash{\hskip.0pt\raise.1pt\hbox{$\Theta$}}\vphantom{}\crcr
      \smash{\hbox{$\Theta$}}\vphantom{$\Theta$}}
      \vphantom{\hbox{$\Theta$}}
      }}
\def\bDelta{{
      \ooalign{
      \smash{\hskip.4pt\raise.4pt\hbox{$\Delta$}}\vphantom{}\crcr
      \smash{\hskip.7pt\raise.6pt\hbox{$\Delta$}}\vphantom{}\crcr
      \smash{\hbox{$\Delta$}}\vphantom{$\Delta$}}
      \vphantom{\hbox{$\Delta$}}
      }}
\def\bLambda{{
      \ooalign{
      \smash{\hskip.5pt\raise.5pt\hbox{$\Lambda$}}\vphantom{}\crcr
      \smash{\hskip.0pt\raise.1pt\hbox{$\Lambda$}}\vphantom{}\crcr
      \smash{\hbox{$\Lambda$}}\vphantom{$\Lambda$}}
      \vphantom{\hbox{$\Lambda$}}
      }}

\makeatletter

\def\bordermatrix#1{\begingroup \m@th
  \@tempdima 8.75\p@
  \setbox\z@\vbox{%
    \def\cr{\crcr\noalign{\kern2\p@\global\let\cr\endline}}%
    \ialign{$##$\hfil\kern2\p@\kern\@tempdima&\thinspace\hfil$##$\hfil
      &&\quad\hfil$##$\hfil\crcr
      \omit\strut\hfil\crcr\noalign{\kern-\baselineskip}%
      #1\crcr\omit\strut\cr}}%
  \setbox\tw@\vbox{\unvcopy\z@\global\setbox\@ne\lastbox}%
  \setbox\tw@\hbox{\unhbox\@ne\unskip\global\setbox\@ne\lastbox}%
  \setbox\tw@\hbox{$\kern\wd\@ne\kern-\@tempdima\left[\kern-\wd\@ne
    \global\setbox\@ne\vbox{\box\@ne\kern2\p@}%
    \vcenter{\kern-\ht\@ne\unvbox\z@\kern-\baselineskip}\,\right]$}%
  \null\;\vbox{\kern\ht\@ne\box\tw@}\endgroup}
\makeatother

\makeatletter
\def\argmin{\mathop{\operator@font arg\,min}}
\def\argmax{\mathop{\operator@font arg\,max}}
\makeatother

\def\bm#1{\mbox{\boldmath $#1$}}

\newcommand{\bea}{\begin{array}}
\newcommand{\ena}{\end{array}}
\newcommand{\beq}{\begin{equation}}
\newcommand{\enq}{\end{equation}}

\newcommand{\beqa}{\begin{eqnarray}}
\newcommand{\enqa}{\end{eqnarray}}

\newcommand{\beqan}{\begin{eqnarray*}}
\newcommand{\enqan}{\end{eqnarray*}}

\newcommand{\AL}{\begin{enumerate}}
\newcommand{\ALE}{\end{enumerate}}

\def\addots{\mathinner{
    \mkern1mu\raise0pt\vbox{\kern7pt\hbox{.}}
    \mkern2mu\raise4pt\hbox{.}
    \mkern2mu\raise7pt\hbox{.}
    \mkern1mu}}

\def\sddots{\mathinner{
    \mkern.8mu\raise7pt\hbox{.}
    \mkern.8mu\raise4pt\hbox{.}
    \mkern.8mu\raise0pt\vbox{\kern7pt\hbox{.}}
    \mkern1mu}}

\def\saddots{\mathinner{
    \mkern.2mu\raise0pt\vbox{\kern7pt\hbox{.}}
    \mkern.2mu\raise4pt\hbox{.}
    \mkern.2mu\raise7pt\hbox{.}
    \mkern1mu}}

\def\sqplus{\mathbin{
	{\ooalign{\hfil\raise.3ex\hbox{\scriptsize
	+}\hfil\crcr\mathhexbox274\crcr\mathhexbox275}}
	}} 
\def\sqminus{\mathbin{
	{\ooalign{\hfil\raise.3ex\hbox{\scriptsize
	--}\hfil\crcr\mathhexbox274\crcr\mathhexbox275}}
	}}

\def\IC{{
   \mathord{
      \hbox to 0em{
	 \hskip-4pt
         \ooalign{
	   \smash{\hskip1.9pt\raise2.6pt\hbox{$\scriptscriptstyle |$}}\crcr
	   \smash{\hbox{\rm\sf C}} }
	 \hidewidth}
      \phantom{\hbox{\rm\sf C}}
} }}
\def\IN{
    {\ooalign{
   \smash{\hskip2.2pt\raise1.5pt\hbox{$\scriptscriptstyle |$}}\vphantom{}\crcr
   \hbox{\sf N}
	}}
	} 
\def\IZ{
    {\ooalign{
   \smash{\hskip1.9pt\raise0pt\hbox{$\sf Z$}}\vphantom{}\crcr
   \hbox{\sf Z}
	}}
	} 
\def\IR{
    {\ooalign{
   \smash{\hskip2.2pt\raise1.5pt\hbox{$\scriptscriptstyle |$}}\vphantom{}\crcr
   \smash{\hskip2.2pt\raise3.3pt\hbox{$\scriptscriptstyle |$}}\vphantom{}\crcr
   \hbox{\sf R}
	}}
	} 

\DeclareMathAlphabet{\mathcmb}{OT1}{cmr}{b}{n}

\def\bSigma{\ensuremath{\mathcmb{\Sigma}}}
\def\bLambda{\ensuremath{\mathcmb{\Lambda}}}

\def\bTheta{\ensuremath{\mathcmb{\Theta}}}

\newcommand{\SI}{\begin{indlist}}
\newcommand{\EI}{\end{indlist}}

\newcommand{\DL}{\begin{dashlist}}
\newcommand{\DLE}{\end{dashlist}}

\makeatletter
\def\setboxz@h{\setbox\z@\hbox}
\def\wdz@{\wd\z@}
\def\boxz@{\box\z@}
\def\underset#1#2{\binrel@{#2}%
  \binrel@@{\mathop{\kern\z@#2}\limits_{#1}}}
\def\binrel@#1{\begingroup
  \setboxz@h{\thinmuskip0mu
    \medmuskip\m@ne mu\thickmuskip\@ne mu
    \setbox\tw@\hbox{$#1\m@th$}\kern-\wd\tw@
    ${}#1{}\m@th$}%
  \edef\@tempa{\endgroup\let\noexpand\binrel@@
    \ifdim\wdz@<\z@ \mathbin
    \else\ifdim\wdz@>\z@ \mathrel
    \else \relax\fi\fi}%
  \@tempa
}
\let\binrel@@\relax%
\makeatother

\usepackage{hyperref}

\begin{document}

\title{Channel Estimation in MIMO Systems with \\One-bit Spatial Sigma-delta ADCs}
\author{R.S. Prasobh Sankar,~\IEEEmembership{Student Member, IEEE}, and Sundeep Prabhakar Chepuri,~\IEEEmembership{Member, IEEE}
\thanks{The authors are with the Department of Electrical Communication Engineering, Indian Institute of Science, Bangalore, India. Email:\{rsprasobh,spchepuri\}@iisc.ac.in. 

This work was supported in part by Nokia Faculty Research Award (NSN Oy, Espoo, Finland) and MHRD, India. The conference precursor of this paper appeared in the 46th IEEE International Conference on Acoustics, Speech and Signal Processing (ICASSP), June 2021~\cite{sankar2021mmWave_SD}. 

}
}

\maketitle

\begin{abstract}
This paper focuses on channel estimation in single-user and multi-user MIMO systems with multi-antenna base stations equipped with 1-bit spatial sigma-delta analog-to-digital converters (ADCs). A careful selection of the quantization voltage level and phase shift used in the feedback loop of 1-bit sigma-delta ADCs is critical to improve its effective resolution. We first develop a quantization noise model for 1-bit spatial sigma-delta ADCs. Using the developed noise model, we then present a two-step channel estimation algorithm to estimate a multipath channel parameterized by the gains, angles of arrival (AoAs), and angles of departure (AoDs). Specifically, in the first step, the AoAs and path gains are estimated using uplink pilots, which excite all the angles uniformly. Next, in the second step, the AoDs are estimated by progressively refining uplink beams through a recursive bisection procedure. For this algorithm, we propose a technique to select the quantization voltage level and phase shift. Through numerical simulations, we demonstrate that with the proposed parametric channel estimation algorithm, MIMO systems with 1-bit spatial sigma-delta ADCs perform significantly better than those with regular 1-bit ADCs and are on par with MIMO systems with high-resolution ADCs.

\end{abstract}

\begin{IEEEkeywords}
Angular channel model, channel estimation, mmWave MIMO, 1-bit quantization, quantization noise modeling, spatial sigma-delta ADC.
\end{IEEEkeywords}

\IEEEpeerreviewmaketitle

\section{Introduction}

\IEEEPARstart{M}{illimeter} wave (mmWave) multiple-input multiple-output (MIMO) systems have become very popular for sensing and wireless communications beyond 5G~\cite{rappaport2013millimeter,alkhateeb2014channel,ghosh2014millimeter}. While the abundant spectrum available at the mmWave frequency bands enables higher cellular data rates and precise positioning, links at mmWave frequencies are very sensitive to blockages and have significantly higher path loss. These issues are alleviated by beamforming with very large antenna arrays, typically packed in small areas. MIMO systems operating at mmWave frequencies, commonly referred to as massive MIMO systems, are either {\it single-user MIMO} (SU-MIMO) systems with multi-antenna user equipment~(UE) and a base station~(BS) having a large antenna array or {\it multi-user MIMO} (MU-MIMO) systems with many single antenna UEs communicating with a BS having a large array.

High-resolution analog-to-digital converters~(ADCs) and digital-to-analog converters~(DACs) for every antenna in the array significantly increase the radio frequency~(RF) complexity and power consumption of massive MIMO systems. Low-resolution quantizers (e.g., 1-bit) are thus preferred albeit their deteriorated performance~\cite{mo2015capacity,li2017channel,rothy2018a_comparison}. Sigma-delta ($\Sigma\Delta$) quantization is a popular technique frequently used to increase the effective resolution of low-resolution quantizers~\cite{aziz1996anoverview}. In a 1-bit $\Sigma\Delta$ quantizer, the time-domain signal is first oversampled at a rate significantly higher than the Nyquist rate. Then the difference between the input and the 1-bit quantized output, i.e., the quantization noise, is fed back in time by adding it to the input at the next time instance. This operation leads to \textit{noise shaping} with the quantization noise pushed to higher temporal frequencies. This means that the effective quantization noise is negligible for a low-pass signal, and it would be as if the signal were quantized by a high-resolution quantizer. This classical architecture to increase the effective resolution of time-domain signals by using a simple 1-bit quantizer with feedback has been recently adapted to the spatial domain~\cite{corey2016spatial,barac2016spatial} and is receiving steady attention for multi-antenna communications~\cite{venkateswaran2011multichannel,shao2019onebit,rao2019massive,rao2020massive,pirzadeh2020spectral}. 

In a {\it 1-bit spatial $\Sigma\Delta$ quantizer}, oversampling and feedback are performed in the spatial domain, i.e., across antennas. To perform spatial oversampling, the antenna elements of an array are placed less than half wavelength apart. The quantization noise of each antenna is fed back along with the input of the next antenna. Analogous to its temporal counterpart, a spatial $\Sigma\Delta$ quantizer pushes and shapes the quantization noise to higher spatial frequencies away from the array broadside. In other words, the quantization noise at lower spatial frequencies in a spatial $\Sigma\Delta$ quantizer is reduced as if it was arising from a higher-resolution quantizer~\cite{corey2016spatial,barac2016spatial}. By introducing phase shifts to the quantization noise before feedback allows {\it angle steering} so that the quantization noise (respectively, the effective resolution) will be lower (respectively, higher) for signals arriving around the steering angle~\cite{shao2019onebit,rao2020massive}. Therefore{\color{black},} with angle steering, it is possible to obtain a higher effective resolution for signals of interest in a spatial sector of certain width centered around any desired angle. In addition, a careful selection of the quantization voltage level assigned to 1-bit quantized signals significantly improves the inference performance when working with spatial $\Sigma\Delta$ quantizers. 

In essence, 1-bit spatial $\Sigma\Delta$ quantization is an attractive architecture for massive MIMO systems, requiring only 1-bit quantizers per antenna element with feedback across the elements. However, feedback and 1-bit quantization make the channel estimation required for beamforming and symbol detection very challenging. This work focuses on channel estimation in MIMO systems with 1-bit spatial $\Sigma\Delta$ quantizers.

\subsection{Related prior works}

For rich scattering environments with a large number of multipath components, the MIMO channel matrix does not have any apparent structure. Such {\it unstructured channel models} are useful for channels at sub-6 GHz frequency bands~\cite{rao2020massive,li2017channel}. In contrast, at mmWave frequencies, due to the extreme path loss, the mmWave channel matrix is sparse in the angular domain and can be parameterized with the angles of departure (AoD), angles of arrival (AoA), and the complex gain of each path. Such {\it angular channel models} are commonly used at mmWave frequencies, e.g., at 28 GHz~\cite{ghosh2014millimeter,alkhateeb2014channel}.

Channel estimation with unstructured models in MIMO systems with 1-bit or few-bit quantizers is typically performed by first linearizing the non-linear quantizer using the so-called \textit{Bussgang decomposition}~\cite{bussgang1952crosscorrelation,demir2020bussgang,li2017channel} followed by computing a linear minimum mean squared error (LMMSE) estimate of the MIMO channel matrix~\cite{li2017channel,jacobsson2017throughput}. Bussgang decomposition based techniques~\cite{li2017channel,demir2020bussgang} of linearizing low-resolution quantizers have also been extended to spatial $\Sigma\Delta$ quantizers for MU-MIMO channel estimation with unstructured models~\cite{rao2019massive,rao2020massive}. 

For channel estimation with {\it angular models}, Bussgang decomposition based methods are not useful as the channel correlation matrix required for computing the Bussgang decomposition, LMMSE estimate, or setting the quantization voltage level in spatial $\Sigma\Delta$ quantizers is not available. This is because knowing the channel correlation matrix for angular channel models amounts to knowing the unknown parameters, namely, AoAs and AoDs that characterize the channel. Thus, for channel estimation in MIMO systems with 1-bit ADCs and angular models, techniques based on optimization to recover the missing amplitudes~\cite{qian2019amplitude}, sparse recovery~\cite{mo2014channel}, and deep learning~\cite{zhang2020DL_massive_MIMO} have been proposed.

To summarize, existing works on channel estimation in MIMO systems with 1-bit spatial $\Sigma\Delta$ quantizers focus on unstructured models~\cite{rao2019massive,rao2020massive}, and they cannot be directly extended to angular channel models. Therefore, in this work, we focus on channel estimation with angular channel models in MIMO systems having 1-bit spatial $\Sigma\Delta$ quantizers.

\subsection{Contributions and main results}
 This paper is an extension of the precursor~\cite{sankar2021mmWave_SD}, wherein we presented a parametric channel estimation technique for SU-MIMO systems with a single line-of-sight~(LoS) path. In this work, we extend~\cite{sankar2021mmWave_SD} in several aspects to estimate multipath channels in SU-MIMO and MU-MIMO systems by leveraging angle steering in spatial $\Sigma\Delta$ quantizers and describe methods to choose the quantization voltage level. The major contributions and results are summarized as follows. 

\begin{itemize}
\item {\it Quantization noise model:} For channel estimation with angular models, as discussed before, Bussgang decomposition based linearization techniques cannot be used. Therefore, we derive a model for the quantization noise in 1-bit spatial $\Sigma\Delta$ quantizers based on the deterministic input-output relation in 1-bit temporal $\Sigma\Delta$ quantizers. Specifically, we derive a closed-form expression for the correlation matrix of the approximation error due to linearization of the 1-bit spatial $\Sigma\Delta$ quantizer. To do so, we use one of the main results of the paper that for most of the antenna elements in a large array, the quantization noise is uncorrelated with the corresponding input. 

\item {\it Channel estimation and quantization voltage selection:} Leveraging the proposed quantization noise model, we develop algorithms to estimate multipath channels admitting angular models for SU-MIMO and MU-MIMO systems. We use {\it uplink pilots} to perform channel estimation at the BS equipped with a 1-bit spatial $\Sigma\Delta$ quantizer. For SU-MIMO and MU-MIMO systems, we present techniques to choose the quantizer voltage level, which, when not chosen carefully, leads to significant performance degradation due to the extreme quantization.

For channel estimation in SU-MIMO systems, i.e., to estimate the AoAs at the BS, AoDs of the paths from the UE, and path gains, we propose a two-step channel estimation algorithm, which is computationally efficient and has low overhead. In \texttt{Step 1}, the multi-antenna UE omnidirectionally transmits pilot symbols to the BS, which estimates the AoAs using a Bartlett beamformer and the complex path gains using a weighted least squares estimator. Since the AoDs are not known in \texttt{Step 1}, the proposed omnidirectional transmission ensures that sufficient power reaches the BS via all the paths. In addition, it allows us to choose a suitable quantization voltage level, which is essential for path gain estimation.
Next, in \texttt{Step~2}, to estimate the AoDs, the UE transmits precoded pilot symbols using a sequence of adaptively chosen beamformers from a codebook hierarchically. To reduce the overhead, we assume a 1-bit feedback link between the BS and UE. We also provide a method to choose a quantization voltage level in \texttt{Step~2}. We show that with angle steering and the proposed voltage level, the beampatterns (from the designed codebook) as seen at the output of the 1-bit spatial $\Sigma\Delta$ quantizer is comparable to that at the input (i.e., without quantization), and thereby resulting in channel estimates that are on par with that of unquantized MIMO systems.
	
We then specialize the SU-MIMO channel estimation algorithm for MU-MIMO systems to estimate the AoAs  and path gains at the BS. Specifically, we reformulate the MU-MIMO channel estimation problem using orthogonal pilots and separately estimate the single-input multiple-output (SIMO) channels between each single antenna UE and multi-antenna BS with a 1-bit spatial $\Sigma\Delta$ quantizer.

\item {\it Performance:} Through numerical simulations, performance of the proposed channel estimation algorithms, in terms of normalized mean squared error (NMSE), are found to be significantly better than that of the regular (non-$\Sigma\Delta$) 1-bit channel estimation algorithms and are comparable with MIMO systems having infinite resolution quantizers for most of the SNRs and for paths with angles not far from the array broadside. Performance of algorithms with 1-bit spatial $\Sigma\Delta$ quantizers is limited for paths with angles away from the array broadside because of the quantization noise shaping. The proposed channel estimation algorithm also performs better than the state-of-the-art unstructured channel estimation algorithm for MIMO systems with 1-bit $\Sigma\Delta$ quantizers~\cite{rao2020massive}, where we use as input to the unstructured channel estimation algorithm a realistic approximation of the channel correlation matrix.  
\end{itemize}

Although we focus on channel estimation with angular models, the developed quantization noise model is useful for estimating unstructured channels using classical estimation techniques (e.g., using least squares) whenever unquantized channel correlation information is not available a priori.

\subsection{Organization and notation}
The remainder of the paper is organized as follows. In Sections~\ref{sec:sigma_delta_review} and~\ref{sec:quant_noise}, we describe  1-bit spatial $\Sigma\Delta$ quantizers and model the quantization noise in 1-bit spatial $\Sigma\Delta$ quantizers, respectively. In Section~\ref{sec:system_model}, we present the SU-MIMO and MU-MIMO system models, which we use for channel estimation. In Sections~\ref{sec:su_mimo} and~\ref{sec:ch_estm_mu_simo}, we propose channel estimation algorithms for SU-MIMO and MU-MIMO systems, respectively. In Section~\ref{sec:simulation}, we discuss results from numerical experiments and conclude the paper in Section~\ref{sec:conclusions}.

Throughout the paper, we use lowercase letters to denote scalars and boldface lowercase (respectively, uppercase) to denote vectors (respectively, matrices). We use $(\cdot)^*$, $(\cdot)\rT$, and $(\cdot)\rH$ to denote complex conjugation, transpose, and Hermitian (i.e., complex conjugate transpose) operations, respectively. $[\vx]_n$ or $x_n$ denotes the $n$-th entry of the vector~$\vx$. $\mA \odot \mB$ denotes the Khatri-Rao (or columnwise Kronecker) product of matrices $\mA$ and $\mB$. Since we restrict ourselves to a 1-bit spatial $\Sigma\Delta$ quantizer only at the BS, henceforth, we simply refer to it as a 1-bit spatial $\Sigma\Delta$ ADC.

Software to reproduce the results in this paper is available at 
\href{https://ece.iisc.ac.in/~spchepuri/sw/SpatialSigmaDelta.zip}{\texttt{\url{https://ece.iisc.ac.in/~spchepuri/sw/SpatialSigmaDelta.zip}}.}

\section{One-bit spatial sigma-delta ADC}
\label{sec:sigma_delta_review}

In this section, we describe the architecture of a multi-channel first-order 1-bit spatial $\Sigma\Delta$ ADC with angle steering. 
Let us denote the input and output of an $N_{\rm r}$ channel 1-bit spatial $\Sigma\Delta$ ADC, at time $t$, as $\vx(t) = [x_1(t),x_2(t),\dots,x_{N_{\rm r}}(t)]\rT \in \mbC^{N_{\rm r}}$ and $\vy(t) = [y_1(t),y_2(t),\dots,y_{N_{\rm r}}(t)]\rT \in \mbC^{N_{\rm r}}$, respectively. 
\begin{figure}[t]
	\centering
	\includegraphics[width=0.8\columnwidth]{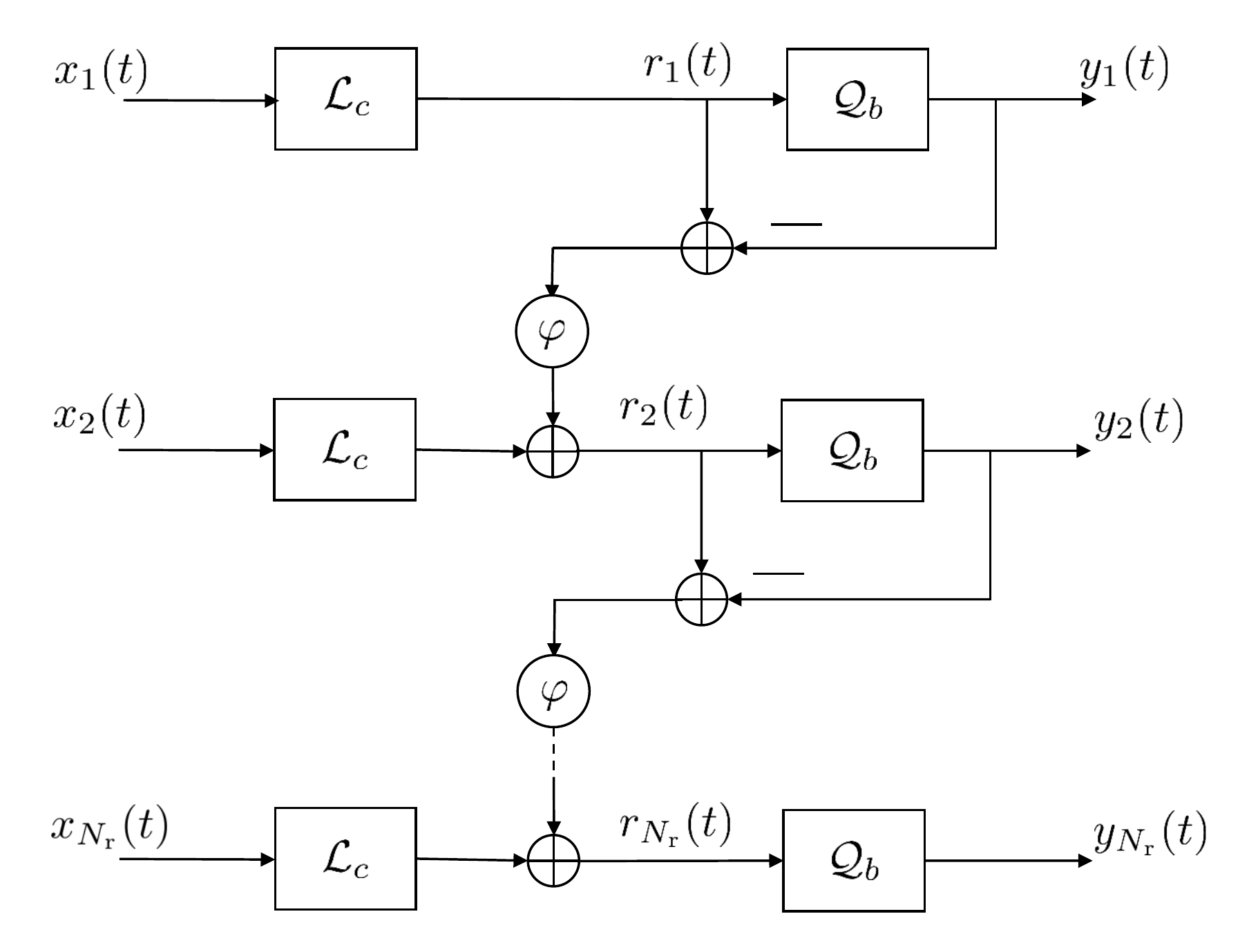}
	\caption{\small The 1-bit spatial $\Sigma\Delta$ ADC architecture. $\mathcal{Q}_b$ is the 1-bit quantizer with level $b$ and $\mathcal{L}_c$ is the amplitude limiter with level~$c$. }
	\label{fig:sig_del_adc}
\end{figure}

Quantization noise from each antenna is fed back along with the input of the next antenna to realize the $\Sigma\Delta$ architecture in space. Due to the presence of the feedback, the quantization noise in a spatial  $\Sigma\Delta$ ADC may become unbounded, overloading the output~\cite{gray1989quantization,shao2019onebit}.  To prevent overloading, the input signal, $x \in \mbC$, is clipped using $\cL_c(\cdot)$ with clipping level $c > 0$ as 
\[
\cL_c[x] = {\rm sign}(\Re(x)) \left\{\vert \Re(x) \vert\right\}_c+ \jmath  \,{\rm sign}(\Im(x)) \left\{\vert \Im(x) \vert\right\}_c,
\]
where the operator $\left\{x\right\}_c$ is defined as $\left\{x\right\}_c = \max\{x,c\}$ and $\jmath = \sqrt{-1}$. The clipped signal is then quantized using a 1-bit quantizer $\cQ_b[\cdot]$ with quantization voltage level $b$ as
\[
\cQ_b[\cL_c[x]] = b\,{\rm sign}(\Re(\cL_c[x])) + \jmath  \,b\,{\rm sign}(\Im(\cL_c[x])).
\]
Let $r_n(t)$ denote the input to {\it channel $n$} of the ADC at time~$t$. Then the corresponding output $y_n(t)$ is given by
\begin{equation} \label{eq:sd_basic_1}
	y_n(t) = \cQ_b[ r_n(t) ] = r_n(t) + e_n(t),
\end{equation}
where $e_n(t) = y_n(t) - r_n(t)$ is the quantization error. The quantization error at {\it channel} $n$ is phase shifted by $\varphi$ and added to the input of {\it channel} $n+1$ as \begin{equation} 
\label{eq:sd_basic_2}
	r_n(t) = \cL_c[x_n(t)] - e^{\jmath{\varphi}} e_{n-1}(t), 
\end{equation}
where $\varphi = 2\pi d \sin{\psi}$ with $\psi$ being the {\it steering angle} and $d$ being the inter-element spacing in the array in wavelengths. Since there is no feedback to the first channel, $e_0(t)=~0$. From \eqref{eq:sd_basic_1} and \eqref{eq:sd_basic_2}, we have the recursion
\begin{align} 
    \label{eq:sd_first}
	y_n(t) &= \cQ_b \left[ \sum_{k=1}^{n} e^{\jmath(n-k){\varphi}} \cL_c[ x_k(t)] \right.\nonumber \\ 
	&\hskip30mm \left.- \sum_{k=1}^{n-1}  e^{\jmath(n-k) {\varphi}} y_k(t)     \right].
\end{align}
We are interested in estimating parameters underlying the input signal $\vx(t)$ from the output of the 1-bit spatial $\Sigma\Delta$ ADC $\vy(t)$. This is a challenging problem because of the cascade of two non-linearities in~\eqref{eq:sd_first}. Moreover, the information loss introduced by the 1-bit spatial $\Sigma\Delta$ ADC is mainly determined by the clipping level $c$ and the quantization level $b$. For a given quantization level $b$, we can prevent the system from overloading by choosing the clipping voltage $c$ as~\cite{shao2019onebit} 
\begin{equation} \label{eq:overload_condition}
	c = b(2 - \vert {\rm cos}(\varphi) \vert - \vert {\rm sin}(\varphi) \vert ).
\end{equation}
A small value of $c$ leads to severe loss in the input information due to clipping, which cannot be compensated later by any choice of the quantization level $b$. Hence, it is necessary to choose a sufficiently large $c$ to avoid information loss due to clipping. In this work, we propose to choose $c$ such that ${\rm Pr}\left(\Vert \cL_c[\vx(t)] - \vx(t) \Vert_2 > \delta \right)  \leq \epsilon $ for appropriately selected constants $\delta,\epsilon >0$. See Section~\ref{sec:su_mimo_voltage} for more details on the selection of $c$. With such a choice of $c$, the output of the clipper can be approximated as $\cL_c[x_n(t)] \approx x_n(t), \> n=1,2,\ldots,N_{\rm r}$.  Then the 1-bit spatial $\Sigma\Delta$ recursion in~\eqref{eq:sd_first} simplifies to 
\begin{equation} \label{eq:sd_vec_model}
	\vy(t) = \cQ_b[\mU \vx(t) - \mV \vy(t)],
\end{equation}
where the $N_{\rm r}\times N_{\rm r}$ lower triangular matrices $\mU$ and $\mV$ are defined as 
\begin{equation*}
	\mU = 		 \begin{bmatrix} 
			1 & & & & \\
			e^{\jmath {\varphi}} & 1 & & & \\
			\vdots & \ddots & \ddots & &\\
			e^{\jmath(N_{\rm r}-1){\varphi}} & \cdots & e^{\jmath{\varphi}} & 1
		\end{bmatrix}
		\quad \text{and} \quad \mV = \mU - \mI 
\end{equation*}
with $\mI$ being the $N_{\rm r} \times N_{\rm r}$ identity matrix. The architecture of the first-order 1-bit spatial $\Sigma\Delta$ ADC is shown in Fig.~\ref{fig:sig_del_adc}.

\section{Quantization noise modeling} \label{sec:quant_noise}
Quantization is a non-linear and irreversible operation that makes its statistical analysis complicated. To simplify the analysis of a quantizer, the usual approach is to linearize the quantizer and account for the error due to linearization through additive noise, which is often assumed to be uniformly distributed and uncorrelated with the input of the quantizer~\cite{gray1987oversampled,gray1989quantization,barac2016spatial,venkateswaran2011multichannel,shao2019onebit}. However, this assumption is reasonable only for a multi-level quantizer with many levels and when the input has a sufficiently large dynamic range.  This classical approach of modeling the approximation error due to linearization as additive uniform noise is not suitable for 1-bit quantizers. Therefore, in what follows, we develop a noise model for 1-bit spatial $\Sigma\Delta$ ADCs. 

To begin with, we first express the input-output relation of a spatial $\Sigma\Delta$ ADC in terms of the ${\rm floor}[\cdot]$ function by drawing inspiration from~\cite{gray1989quantization}, where a similar expression for the deterministic error in a temporal 1-bit $\Sigma\Delta$ quantizer was developed. Next, we propose to linearize the ${\rm floor}[\cdot]$ function by interpreting it as a multi-level quantizer to compute the second-order statistics of the quantization noise that is useful for decorrelating observations when solving parametric estimation and detection problems involving 1-bit spatial~$\Sigma\Delta$~ADCs.

The error  $e_n(t)$ in~\eqref{eq:sd_basic_1} due to 1-bit spatial $\Sigma\Delta$ quantization admits a closed-form expression as given in the next Lemma.
\begin{lemma}
\label{theo:error}
For a  1-bit spatial $\Sigma\Delta$ ADC with $\psi = 0$, the quantization error as a function of the input is given by
\begin{subequations}	
\label{eq:sd_realimag}
	\begin{eqnarray} 
	&\Re( {e_{n}(t)}) = b - 2b  \left\langle  \frac{1}{2}(n-1) + \frac{1}{2b} \sum\limits_{k=1}^{n}\Re( x_{k}(t) ) \right\rangle \label{eq:sd_real1} \\
 	&\Im( {e_{n}(t)}) = b - 2b \left\langle \frac{1}{2}(n-1) +\frac{1}{2b} \sum\limits_{k=1}^{n} \Im(x_{k}(t))\right\rangle  \label{eq:sd_imag1}
\end{eqnarray}
\end{subequations}
for $n = 1,\ldots,N_{\rm{r}}$. Here, $\langle\cdot\rangle$ is the fractional part function.
\end{lemma}
The above expressions are derived in the appendix. Let us collect the output of all the channels of the 1-bit spatial $\Sigma\Delta$ ADC in a vector $\vy(t)$ and rewrite~\eqref{eq:sd_vec_model} as
\begin{equation} \label{eq:sd_linear_vec_new}
	\vy(t) = \mU \vx(t) - \mV \vy(t) + \ve(t).
\end{equation}
From Lemma~\ref{theo:error}, we have
\begin{equation} \label{eq:e_t_def}
	\ve(t) = 2b\mu \boldsymbol{1} - 2b \left\langle {\boldsymbol \nu} - \frac{1}{2b}\mU \vx(t)  \right\rangle,
\end{equation}
where ${\boldsymbol \nu} = \mu \mV \boldsymbol{1} \in \mbC^{N_{\rm r}}$ with $\mu= 0.5 + \jmath0.5$, $\boldsymbol{1}$ is the length-$N_{\rm r}$ column vector with all ones, and the fractional part function is applied elementwise. Using the fact that $\langle x \rangle = x - {\rm floor}[x], \> \forall \> x \in \mathbb{R}$, and from~\eqref{eq:sd_linear_vec_new}, the {\it deterministic} relation between the input and output of a 1-bit spatial $\Sigma\Delta$ ADC is given by
\begin{equation} \label{eq:sd_floor_v1}
	 0.5b^{-1}\mU \vy(t) + {\boldsymbol \nu} - \mu\boldsymbol{1} = {\rm floor}\left[  0.5b^{-1}\mU\vx(t) + {\boldsymbol \nu}  \right],
\end{equation}
where we have transformed the non-linearity due to the 1-bit quantization $\cQ_b[\cdot]$ to the non-linear ${\rm  floor[\cdot]}$ function.
\begin{figure}[t]
\centering
\includegraphics[width = 0.75\columnwidth]{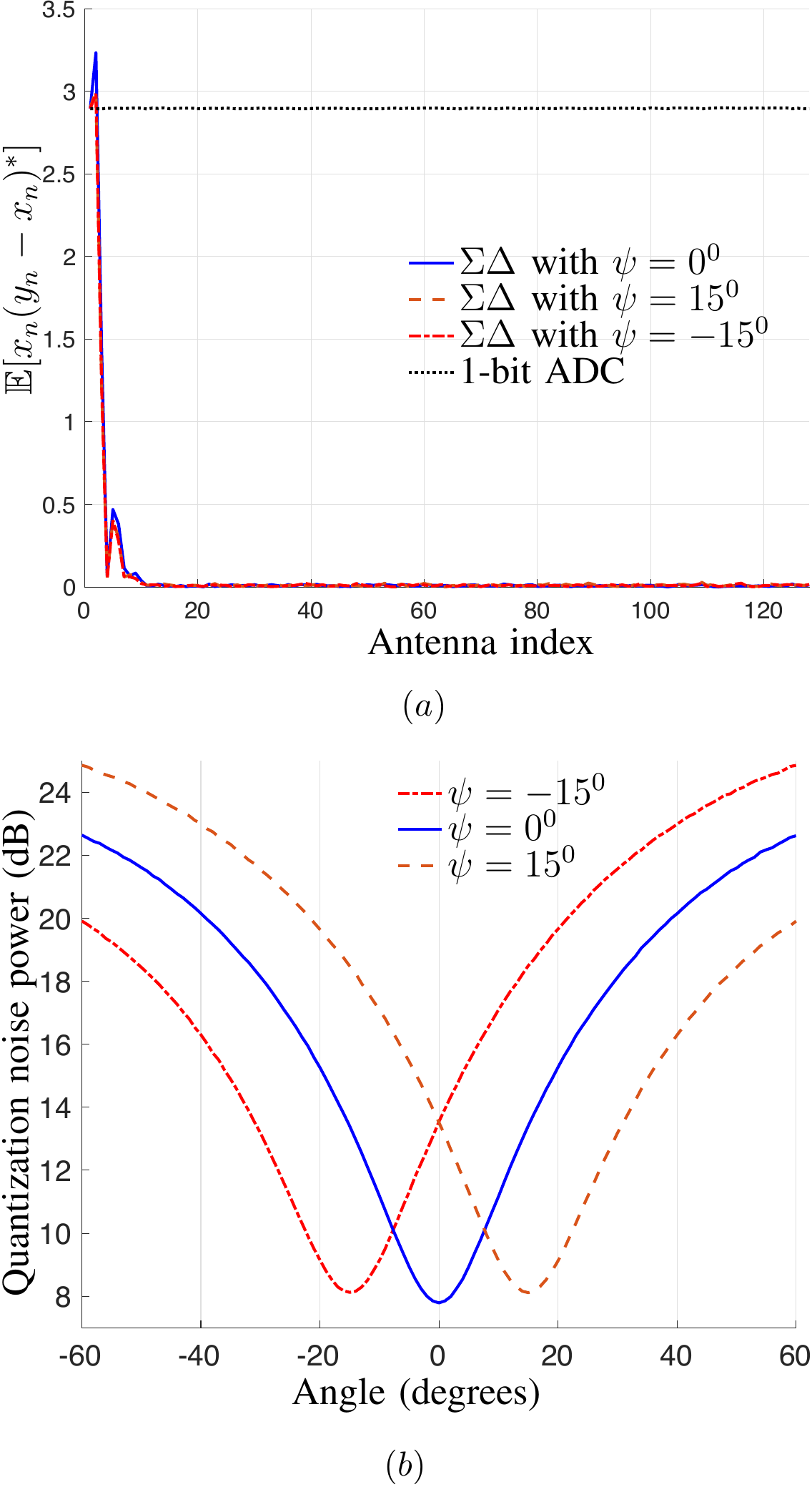}
	\caption{\small Quantization noise in a 128 channel 1-bit $\Sigma\Delta$ ADC\protect\footnotemark. (a)~Correlation between the input and quantization noise of a regular one-bit and a one-bit spatial $\Sigma\Delta$ ADC for different steering angles. (b) Quantization noise shaping.}
	\label{fig:sd_noise}
\end{figure}
\footnotetext{In Fig.~\ref{fig:sd_noise}, the correlation and noise power spectrum are computed by averaging over $10^4$ independent experiments in which we use the same simulation setting as in Fig.~\ref{fig:sumimo_l1} in Section~\ref{sec:simulation:sumimo} with a path impinging on the BS array from the angular sector $[-30^\circ,30^\circ]$ and with a signal-to-noise ratio of 0 dB.}

Now we leverage the fact that the input-output relation of a ${\rm floor}[\cdot]$ function is similar to that of a multi-level quantizer with a step size of one. For the spatial $\Sigma\Delta$ ADC channels (corresponding to antenna elements) with larger indices that are away from the $1$st channel, the dynamic range of the real and imaginary parts of $[0.5b^{-1}\mU\vx(t) + {\boldsymbol \nu}]_n$ is large when compared to $[0,1)$. This means that for antenna elements with indices away from the $1$st element, the ${\rm floor}[\cdot]$ function acts like a \emph{multi-level quantizer} with an input having a large dynamic range. Therefore, it is now reasonable to model the approximation error due to the linearization of the ${\rm floor}[\cdot]$ function as additive noise, $\vw(t) \in~\mathbb{C}^{N_{\rm r}}$, which is uniformly distributed and uncorrelated with the input for channels having larger indices, i.e., we have
\begin{equation*} \label{eq:floor_linear}
{\rm floor}\left[ 0.5b^{-1}\mU\vx(t) + {\boldsymbol \nu}  \right] = 0.5b^{-1}\mU\vx(t) + {\boldsymbol \nu}   + \vw(t)
\end{equation*}
with the real and imaginary parts of $[\vw(t)]_n \in [0,1)$. Substituting in \eqref{eq:sd_floor_v1} yields
\begin{equation} \label{sd_final}
	\mathbf{y}(t) = \mathbf{x}(t) +  \vq(t),
\end{equation}
where $\vq(t) = 2b \mU^{-1} ({\vw}(t) + \mu\boldsymbol{1})$, which is an affine transformation of $\vw(t)$, is also uniformly distributed and $[\vq(t)]_n$ is uncorrelated with the input $[\vx(t)]_n$ for larger channel (or antenna) indices $n$. In massive MIMO systems with large number of antennas, there are many antennas for which $x_n(t)$ and $[\vq(t)]_n$ are uncorrelated. The correlation of ${[\vq(t)]}_n = y_n(t) - x_n(t)$ with $x_{n}(t)$, i.e., $\mathbb{E}[x_n(y_n - x_n)^*]$ for an antenna array having $N_{\rm r} = 128$ elements with an inter-element spacing of one eighth the signal wavelength is illustrated in~Fig.~\ref{fig:sd_noise}(a).

The covariance matrix of the noise vector $\vq(t)$ in~\eqref{sd_final} is given by $\mR_q = \mbE\left[\vq(t) \vq\rH(t) \right] = \frac{2b^2}{3}\mU^{-1}\mU\rHm$. When $\psi = 0$, the matrix $\mU^{-1}$ with ones on the main diagonal, $-1$ on the first sub-diagonal, and zeros elsewhere, is a spatial high-pass filter, which shapes the quantization noise to higher spatial frequencies. We illustrate the angular power spectrum of the quantization noise, i.e., $\mathbb{E}[\vert \va_{\rm BS}\rH(\theta)(\vy(t) - \vx(t)) \vert^2]$, for different values of the steering angle in Fig.~\ref{fig:sd_noise}(b), where $\va_{\rm BS}(\theta) = [1 , e^{-\jmath2\pi d {\rm sin}(\theta)}, \ldots, e^{-\jmath(N_{\rm r}-1) 2 \pi d {\rm sin}(\theta)} ]\rT$ is the steering vector of the uniform linear array (ULA) having $N_{\rm r}$ elements at the BS with an inter-element spacing $d$ wavelengths. We can see that the quantization noise is very small for the directions around the steering angle. 

Before we end this section, we make the following two remarks. Although in the development of the noise model, the spatial $\Sigma\Delta$ ADC is steered to the array broadside with $\psi = 0$ for ease of exposition, introducing a phase shift in the feedback does not alter the uncorrelatedness between the input and the quantization noise as can be seen in Fig.~\ref{fig:sd_noise}(a). Next, for the ${\rm  floor}[\cdot]$ function in~\eqref{eq:sd_floor_v1} to behave as a multi-level quantizer, the quantization voltage level $b$ plays an important role. When $b$ is very large, the dynamic range of the real and imaginary parts of the argument of the ${\rm  floor}[\cdot]$ function, i.e., $0.5b^{-1}\mathbf{U}\mathbf{x}(t) + {\boldsymbol \nu}$, can be very small for which the uniform noise or uncorrelated noise assumption might fail. Hence, carefully selecting $b$ (as discussed in Section~\ref{sec:su_mimo_voltage}) becomes crucial.

Next, we use the developed spatial $\Sigma\Delta$ signal and noise models to estimate MIMO channels with angular models.

\section{Angular channel model}\label{sec:system_model}
In this paper, we consider the two commonly encountered SU-MIMO and MU-MIMO settings in  MIMO communications. In the SU-MIMO setting, a single UE with $N_{\rm t}$ antennas communicates with a multi-antenna BS, whereas in the MU-MIMO setting, $K$ single antenna UEs communicate with a multi-antenna BS. In both these settings, the BS has a ULA with $N_{\rm r}$ antennas and it receives and processes uplink training pilots. 

Let us collect the uplink training pilots in $\mS = [ \vs(1), \vs(2),  \ldots,  \vs(T)]$, where $T$ is the pilot length. Without loss of generality, let us assume that the columns of the pilot matrix have unit norm. Let $P$ denote the total transmit power at the UE and $\mH$ denote the MIMO channel matrix. The signal received at the BS prior to quantization, denoted as $\mX = [\vx(1), \vx(2), \ldots, \vx(T)] \in \mbC^{N_{\rm r} \times T}$, can be expressed as
\begin{equation} \label{eq:mu_simo_model}
	\mX = \sqrt{P}\mH \mS + \mZ,
\end{equation}
where $P$  is also the uplink SNR of the system and $\mZ \in \mbC^{N_{\rm r} \times T}$ is the additive white Gaussian receiver noise matrix with entries $[\mZ]_{ij} \sim \cC \cN (0,1)$. The received signal is then quantized using an $N_{\rm r}$-channel 1-bit spatial $\Sigma\Delta$ ADC to obtain $\mY = [\vy(1), \vy(2), \ldots, \vy(T)] \in \mbC^{N_{\rm r} \times T}$, which is given by
\begin{align*}
	\mY = \cQ_b [\mU \mX - \mV \mY] \nonumber = \cQ_b [\sqrt{P} \mU \mH \mS - \mV \mY + \mU\mZ ].	
\end{align*}
 
Next, we parameterize $\mH$ by assuming a narrowband spatially sparse model for MU-MIMO and SU-MIMO systems.

\subsection{SU-MIMO channel model} \label{sec:su_mimo_channel_model}

Suppose there are $L$ scatterers that result in $L$ paths between the UE and  BS. Let us denote the AoDs of the $L$ paths from the UE by
$\boldsymbol{\phi} = [ \phi_1,\ldots,\phi_L]\rT$, the AoAs of the $L$ paths at the BS by $\boldsymbol{\theta} = [\theta_1,\ldots,\theta_L]\rT$, and the complex path gains by $\boldsymbol{\alpha} = [ \alpha_1,\ldots,\alpha_L]\rT$. The MIMO channel matrix is then expressed in terms of these parameters as 
\begin{align}
	\mH &= \frac{1}{\sqrt{L}}\sum_{k=1}^{L} \alpha_k \va_{\rm BS}(\theta_k) \va_{\rm UE}\rH(\phi_k) \nonumber \\
	&= \frac{1}{\sqrt{L}}\mA_{\rm BS}(\boldsymbol{\theta}) {\rm diag}(\boldsymbol{\alpha}) \mA_{\rm UE} \rH (\boldsymbol{\phi}), \label{eq:channel_sumimo}
\end{align}
where $\mA_{\rm BS}(\boldsymbol{\theta}) = [
	\va_{\rm BS}(\theta_1) , \ldots , \va_{\rm BS}(\theta_L)] \in \mbC^{N_{\rm r} \times L}$ is the array manifold of the ULA at the BS and 
$\mA_{\rm UE}(\boldsymbol{\phi}) = [\va_{\rm UE}(\phi_1) , \ldots , \va_{\rm UE}(\phi_L) ] \in \mbC^{N_{\rm t} \times L}$ is the  array manifold at the UE. The columns of $\mA_{\rm UE}(\boldsymbol{\phi})$ contain the array response vector of the critically spaced ULA at the UE with $N_{\rm t}$ elements, and is given by
$ \va_{\rm UE}(\phi) = [1 , e^{-\jmath\pi {\rm sin}(\phi)} , \ldots , e^{-\jmath(N_{\rm t}-1) \pi {\rm sin}(\phi)}]\rT $.
Similarly, the columns of $\mA_{\rm BS}(\boldsymbol{\theta})$ contain the array response vector of the \textit{oversampled} ULA at the BS with an inter-element spacing $d$ wavelengths, and is given by $ \va_{\rm BS}(\theta) = [1 , e^{-\jmath2\pi d {\rm sin}(\theta)} , \ldots , e^{-\jmath(N_{\rm r}-1) 2 \pi d {\rm sin}(\theta)} ]\rT$.

\subsection{MU-MIMO channel model} \label{subsec:mu_simo}
Suppose that there are $K$ single antenna users. The SIMO channel $\vh_k \in \mbC^{N_{\rm r}}$ between the $k$th UE and the BS with $L_k$ paths is given by
\begin{equation} 
\label{eq:mumimo_channel}
	\vh_k = \frac{1}{\sqrt{L_k}}\sum_{j=1}^{L_k}\alpha_{k,j}\va_{\rm BS}(\theta_{k,j}) = \frac{1}{\sqrt{L_k}}\mA_{\rm BS}(\boldsymbol{\theta}_{k}) \boldsymbol{\alpha}_k,
\end{equation}
where $\mA_{\rm BS}(\boldsymbol{\theta}_{k}) = [
	\va_{\rm BS}(\theta_{k,1}) , \ldots , \va_{\rm BS}(\theta_{k,L_k})
] \in \mbC^{N_{\rm r}\times L_k}$ denotes the array manifold at the BS, $\boldsymbol{\theta}_k = [
	\theta_{k,1} , \ldots, \theta_{k,L_k}
]\rT$ collects AoAs of paths from $k$th UE at the BS, and $\boldsymbol{\alpha}_k = [	\alpha_{k,1} , \ldots , \alpha_{k,L_k}]\rT \in \mbC^{L_k}$ denotes the corresponding complex path gains. Then the overall  channel matrix for the MU-MIMO system is given by
\begin{equation*} 
	\mH = \begin{bmatrix}
		\vh_1 & \vh_2 & \ldots & \vh_K
	\end{bmatrix} \in \mbC^{N_{\rm r} \times K}.
\end{equation*}

Our aim is to estimate the MIMO channels~\eqref{eq:channel_sumimo} and~\eqref{eq:mumimo_channel} by estimating the underlying angles (namely, the AoAs and AODs) and path gains from the output of the 1-bit spatial $\Sigma\Delta$ ADC at the BS given the uplink pilots $\mS$, the SNR $P$, and the number of paths. 

\section{SU-MIMO channel estimation} \label{sec:su_mimo}
In this section, we present the proposed algorithm for channel estimation with angular models in SU-MIMO systems having 1-bit spatial $\Sigma\Delta$  ADCs at the BS. Specifically, we propose a two-step algorithm to estimate the channel parameters $(\boldsymbol{\theta},\boldsymbol{\phi},\boldsymbol{\alpha})$ from uplink pilots. In the first step, we estimate the AoAs (i.e., $\boldsymbol{\theta}$) and the path gains (i.e., $\boldsymbol{\alpha}$) using precoded uplink pilots, which excite all the angles uniformly. Next, in the second step, to estimate the AoDs, i.e., $\boldsymbol{\phi}$, we select precoders from a codebook using a recursive bisection procedure that leverages 1-bit feedback between the BS and UE to reduce the number of channel uses and thus the channel estimation overhead. 

As we discuss later, it is crucial to carefully select the clipping voltage levels to benefit from the advantages of the 1-bit spatial $\Sigma\Delta$ ADCs. The proposed two-step channel estimation procedure is developed keeping in mind the dependence of the unknown parameters on voltage level selection.

\subsection{\texttt{Step 1}: AoA and path gain estimation}  \label{sec:su_mimo_p1}

\subsubsection{AoA estimation} To estimate the AoAs at the BS, the UE transmits precoded pilot symbols $\vs(t) = \vp_1(t)$ for $t=1,\ldots,T_1$, where $\vp_1(t) \in \mbC^{N_{\rm t}}$ is the precoder with $\Vert \vp_1(t) \Vert_2=1$ so that the total transmit power is $P$. As we do not yet know the AoDs, instead of selecting $\vp_1(t)$ to focus energy in specific directions, we select it such that all the departure angles are excited uniformly as
\begin{equation} \label{eq:condition_for_f1}
	\vp_1\rH(t) \va_{\rm UE}(\tilde{\phi}_d) = 1, \, d=1,2,\ldots,D,
\end{equation}
where $\cD = \{ \tilde{\phi}_1, \tilde{\phi}_2, \ldots, \tilde{\phi}_{D}\}$ denotes the set of $D$ candidate AoDs. 
An obvious choice of $\vp_1(t)$ that satisfies \eqref{eq:condition_for_f1} is 
\begin{equation} \label{su_mimo_aoa_precoder}
	\vp_1(t) = \vp_1 = \begin{bmatrix} 1 & 0 & \ldots & 0 \end{bmatrix}\rT, \,\, t=1,\ldots,T_1.
\end{equation}
This means that we turn off all the antennas at the UE except the first one to perform an omnidirectional transmission. 

From~\eqref{sd_final}, the symbols received at the BS in~\texttt{Step 1} can be compactly expressed as
\begin{align}
	\mY_1 &= \sqrt{\frac{P}{L}} \mA_{\rm BS}(\boldsymbol{\theta}) {\rm diag}(\boldsymbol{\alpha}) \mA_{\rm UE}\rH (\boldsymbol{\phi}) \mS + \mN \nonumber \\
	&= \sqrt{\frac{P}{L}} \mA_{\rm BS}(\boldsymbol{\theta}) {\rm diag}(\boldsymbol{\alpha}) \mE + \mN, \label{sumimo:aoa_estm_received_signal}
\end{align}
where $\mS = \vp_1{\bf 1}\rT \in \mbC^{N_{\rm t} \times T_1}$ is the transmitted pilot matrix and the effective noise matrix $\mN = [\vn(1),\cdots,\vn(T_1)] \in \mbC^{N_{\rm r} \times T_1}$ is defined as $\mN= \mZ + \mQ$ is the sum of the receiver noise $\mZ$ and the quantization noise $\mQ = [\vq(1), \cdots, \vq(T_1)]$. The $L \times T_1$ matrix $\mE {=} \mA_{\rm UE}\rH (\boldsymbol{\phi}) \mS = \mA_{\rm UE}\rH (\boldsymbol{\phi}) \vp_1{\bf 1}\rT\in \mbC^{L \times T_1}$ has all one entries. Thus, the precoder in~\eqref{su_mimo_aoa_precoder} makes $\mY_1$ independent of the AoDs.

The AoAs can now be estimated from $\mY_1$ using standard direction-finding techniques like Bartlett beamforming, minimum variance distortionless response~(MVDR) beamforming~\cite{vantrees2002optimum}, sparse recovery, or using a maximum likelihood based estimator. 
Since a massive MIMO BS typically has a large number (of the order of hundreds) of antennas, we may use a computationally less intensive method, such as the Bartlett beamforming for AoA estimation. Specifically, the AoA estimates, denoted by $\{ \hat{\theta}_k\}_{k=1}^{L}$, are the locations of the $L$ local maxima of the Bartlett spatial spectrum
\begin{equation} \label{su_mimo_aoa_spectrum}
	\mathcal{J}(\theta) = \va_{\rm BS}\rH  (\theta) \mR_{y1} \va_{\rm BS}(\theta),
\end{equation}
where $\mR_{y1} = \frac{1}{T_1}\mY_1\mY_1\rH$ is the sample covariance matrix. 

\subsubsection{Path gain estimation} Next, to estimate the path gains, we form the estimated array manifold at the BS as $\hat{\mA}_{\rm BS} = [\va_{\rm BS}(\hat{\theta}_1), \va_{\rm BS}(\hat{\theta}_2), \cdots,	\va_{\rm BS}(\hat{\theta}_L)]$
and approximate~\eqref{sumimo:aoa_estm_received_signal} as
\begin{equation} \label{eq:rec_signal_aoa}
	\hat{\mY}_1  = \sqrt{\frac{P}{L}} \hat{\mA}_{\rm BS} {\rm diag}(\boldsymbol{\alpha}) \mE + \mN.
\end{equation}

Recall from the quantization noise modeling in Section~\ref{sec:quant_noise} that the covariance matrix of the quantization noise is given by $\mR_q = \frac{2b^2}{3}\mU^{-1}\mU\rHm$ and that the quantization noise is uncorrelated with the input. Thus the covariance matrix of the effective noise  term in \eqref{sumimo:aoa_estm_received_signal} is
 \begin{equation} \label{eq:eq_noise_covariance}
	\mR_{n} = \mathbb{E} [\vn(t)\vn\rH(t)] = \mathbf{I} + \frac{2b^2}{3}\mathbf{U}^{-1}\mathbf{U}\rHm.
 \end{equation}
We now prewhiten the observations $\hat{\mY}_1$ to obtain 
\[
{\mY}_1' = \mR_{n}^{-1/2} \hat{\mY}_1 = \sqrt{\frac{P}{L}}\mR_{n}^{-1/2} \hat{\mA}_{\rm BS} {\rm diag}(\boldsymbol{\alpha}) \mE + \mN',
\]
where $\mR_{n}^{-1/2}$ is the prewhitening matrix, which can be obtained using an eigenvalue decomposition of $\mR_{n}$ and $\mN' = \mR_n^{-1/2}\mN$ is the whitened noise term. Using the property that ${\rm vec}(\mA {\rm{diag}(\vb)}\mC) = (\mC\rT \odot \mA)\vb$, we have ${\rm vec}({\mY}_1') = \boldsymbol{ \Psi } \boldsymbol{\alpha} + {\rm vec}(\mN')$ with $\boldsymbol{ \Psi } = ( \mE \rT \odot  \sqrt{\frac{P}{L}} \mR_n^{-1/2} \hat{\mA}_{\rm BS}  ) \in \mbC^{T_1 N_{\rm r} \times L}$ and $T_1N_{\rm r} \gg L$ as the BS usually has large number of antennas and the mmWave MIMO channel is sparse in the angular domain. The path gains can then be estimated using least squares as 
\begin{equation} \label{eq:su_alpha_est}
	\hat{\boldsymbol{\alpha}} = (\boldsymbol{\Psi}\rH\boldsymbol{\Psi})^{-1}\boldsymbol{\Psi}\rH {\rm vec}({\mY}_1').	
\end{equation}

\subsection{\texttt{Step 2}: AoD estimation}  \label{sec:su_mimo_p2}

Now what remains is to estimate the AoDs. To do so, we propose a recursive bisection procedure, which divides the spatial sector into two subsectors at each stage and measures the power of the signal received in the direction corresponding to the estimated AoA. The method selects the subsector with the largest received power as the new sector to be used in the next bisection stage. This procedure is continued till the desired subsector resolution is obtained and is repeated for each of the estimated AoAs corresponding to $L$ paths. 

Let us denote the precoder that we use in \texttt{Step 2} by $\vp$ with $\Vert \vp \Vert = 1$, as before. Let us also define the inner product between $\va_{\rm UE}(\phi_l)$ and $\vp$ as $\rho_l(\vp) = \va_{\rm UE}\rH (\phi_l)\vp$, where we have $0 \leq \vert \rho_l(\vp) \vert \leq \sqrt{N_{\rm t}}$ and when $\vp = \frac{1}{\sqrt{N_{\rm t}}}\va_{\rm UE}(\phi_l)$, the upper bound is achieved with equality. Therefore, by choosing a precoder $\vp$ that yields the maximum $\rho_l(\vp)$, we can indirectly refine the AoD sector of the $l$th path to compute the AoD.

Suppose the UE transmits $T_2$ precoded symbols $\mS = \vp{\bf 1}\rT \in \mathbb{C}^{N_{\rm t} \times T_2}$, where the subscript ``2" denotes \texttt{Step 2} of the algorithm. Using the combiner $\vc_l = \frac{1}{\sqrt{N_{\rm r}}} \va_{\rm BS}({\theta}_l)$, the component of the received uplink pilot arriving along the $l$th path at the BS can be expressed as 
\begin{align}
	\vc_l\rH \mY &= \sqrt{\frac{P}{L}}\sum_{k=1}^{L}\alpha_k \vc_l\rH\va_{\rm BS}(\theta_k) \va_{\rm UE}\rH (\phi_k)\vp\boldsymbol{1}\rT + \vc_l\rH\mN \nonumber \\
	&= \sqrt{\frac{P}{L}} \alpha_l \vc_l\rH \va_{\rm BS}(\theta_l) \rho_l(\vp)\boldsymbol{1}\rT +   \vc_l\rH\mN  \nonumber \\
	& \quad \quad \quad + \sqrt{\frac{P}{L}} \sum_{k=1,k \neq l}^{L}	\alpha_k \vc_l\rH \va_{\rm BS}(\theta_k) \rho_k(\vp)\boldsymbol{1}\rT.
\label{eq:su_mimo_doa_est_idea}
\end{align}
Since we have a large array at the BS and assuming that the AoAs are sufficiently separated, we can approximate $\vc_l\rH \va_{\rm BS}(\theta_k) = \frac{1}{\sqrt{N_{\rm r}}}\va_{\rm BS}\rH({\theta}_l)\va_{\rm BS}(\theta_{k}) \approx 0$ for $l \neq k$. Using this approximation and multiplying both sides of~\eqref{eq:su_mimo_doa_est_idea} with ${\bf 1}$, we have
\begin{equation} \label{eq:rec_signal_after_bf}
	\vc_l\rH\mY{\bf 1} = \sqrt{\frac{PN_{\rm r}}{L}}T_2\alpha_l \rho_l(\vp) + \vc_l\rH\mN{\bf 1},
\end{equation}
which allows us to compute the energy of the $T_2$ received symbols arriving from the $l$th path as
\begin{equation}
E_{l}(\vp) = \left|\frac{1}{T_2}\vc_l\rH\mY{\bf 1}\right|^2.
\label{eq:precoder_energy}
\end{equation}
In practice, we compute $E_l({\vp})$ using the estimate $\hat{\theta}_l$ from \texttt{Step 1} to form the combiner as ${\vc}_l = \frac{1}{\sqrt{N_{\rm r}}}\va_{\rm BS}(\hat{\theta}_l)$. 
\begin{figure*}[t]
\includegraphics[width = 2\columnwidth]{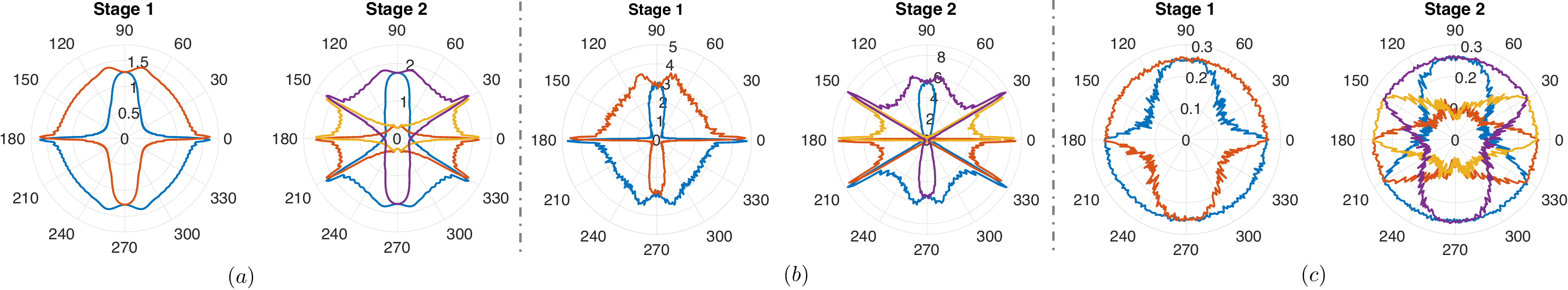}
	\caption{\small (a) First two stages of the precoders that partition the spatial sector into two and four subsectors. Corresponding beampatterns as seen at the BS (b) without any quantizer and (c) with 1-bit $\Sigma\Delta$ ADC, but without any angle steering or voltage level selection. The UE has $N_{\rm t} = 32$ critically-spaced antennas and the BS has $N_{\rm r} = 128$ antennas spaced one eighth wavelengths apart.}
	\label{fig:sd_precoders}
\end{figure*}

We can estimate the AoDs by finding precoders, $\vp$, that maximize $E_l({\vp})$ for each path $l=1,\ldots,L$, where recall that the structure of the precoder vector is constrained to $\vp = \frac{1}{\sqrt{N_{\rm t}}}\va_{\rm UE}(\phi)$. Here, we assume that the number of paths, $L$, is known. Performing this maximization by exhaustively searching over all the departure angles in a predefined grid ${\cD}$ results in excessive training overhead. Therefore, we present a recursive bisection procedure, where for each path $l$, we start with a wide beam and progressively refine it to maximize~$E_l({\vp})$. 

\subsubsection{Design of the precoder codebook} \label{sec:su_mimo_dod_codebook}
Before we describe the proposed recursive biscection procedure, we first provide the design of a codebook that contains precoders required for each stage of the procedure. Let us discretize the departure angular domain into $D$ grid points $\cD = \{ \tilde{\phi}_1, \tilde{\phi}_2, \cdots, \tilde{\phi}_{D}\}$ by uniformly sampling the direction cosine space so that $\tilde{\phi}_d = {\rm sin}^{-1}\left(-1 + \frac{2}{D-1}(d-1)\right)$ for $d=1,2,\ldots,D$.  Let us define the index set $\cI_{s,i}$  as
\begin{equation*}
	\cI_{s,i} = \left\{ \frac{D(i-1)}{2^s} + 1, \frac{D(i-1)}{2^s} + 2, \ldots, \frac{Di}{2^s}\right\}
\end{equation*}
for $i=1,2,\ldots,2^s$ with $\vert \cI_{s,i} \vert = \frac{D}{2^s}$. Next, we the partition the set $\cD$ into $2^s$ partitions for the $s$th stage of the recursive procedure with a total number of $N_{\rm s} = {\rm log}_2(D)$ stages. In other words, at Stage $s$, we have $2^s$ spatial sectors with the $i$th sector formed from angles $\tilde{\phi}_i \in \cI_{s,i}$ and in the last stage, i.e., at Stage $N_{\rm s}$, we have sectors with $D$ angular grid points in $\mathcal{D}$. For example, in the first stage, we have two spatial sectors $\{\tilde{\phi}_1,\tilde{\phi}_2,\cdots,\tilde{\phi}_{\frac{D}{2}}\}$ and $\{\tilde{\phi}_{\frac{D}{2}+1}, \tilde{\phi}_{\frac{D}{2}+2},\cdots, \tilde{\phi}_D\}$. 

Let $\vp_{s,i}$ denote the $i$th precoder for the $s$th stage. The precoders are designed to focus on a desired angular sector as
 \begin{equation} \label{eq:precoder_condition}
	\va_{\rm UE}\rH(\tilde{\phi}_j)\vp_{s,i} = \begin{cases}
		1, & \quad j \in \cI_{s,i}, \\
		0, & \quad \text{otherwise}
	\end{cases}
\end{equation}
for $s = 1,2,\ldots,N_{\rm s}$ and $i=1,2,\ldots,2^s$. Defining the dictionary matrix $\mD = [\va_{\rm UE}(\tilde{\phi}_1), \ldots , \va_{\rm UE}(\tilde{\phi}_{D})]\rH \in \mbC^{ D \times N_{\rm t}}$ with $D \gg N_{\rm t}$, and the precoder matrix $\mP_s = [\vp_{s,1} , \vp_{s,2}, \ldots, \vp_{s,2^s}]~\in~\mathbb{C}^{N_{\rm t} \times 2^s}$, we can compactly express~\eqref{eq:precoder_condition} as the linear system
\begin{equation}
	\mD  \mP_s = \mPsi_s,
\end{equation}	
where the desired beampattern matrix $\mPsi_s = [\mathbbm{1}_{s,1},\cdots, \mathbbm{1}_{s,2^s}] 	\in \{1,0\}^{D \times 2^s}$ with $\mathbbm{1}_{s,i} \in \{0,1\}^{D}$ being the indicator vector with entries equal to one at locations indexed by the set $\cI_{s,i}$. Then the precoders for the $s$th stage can be computed using least squares as
\begin{equation} \label{su_mimo_h_codebook}
	\mP_s = (\mD\rH \mD )^{-1} \mD\rH \mPsi_s.
\end{equation} 
To obtain unit-norm precoding vectors, we normalize each column of $\mP_s$ to unity. We repeat this procedure for $s=1,2, \ldots, N_{\rm s}$ to compute precoders for all the stages. This completes the design of the precoder codebook. 

In Fig.~\ref{fig:sd_precoders}(a), we show the beampattern $|\va_{\rm UE}\rH (\phi)\vp|^2$ for $\phi \in [0,2\pi]$  corresponding to Stage 1 with $\vp \in \{\vp_{1,1},\vp_{1,2}\}$ and Stage 2 with $\vp \in \{\vp_{2,1},\vp_{2,2},\vp_{2,3}, \vp_{2,4}\}$ of the codebook. In Figs.~\ref{fig:sd_precoders}(b) and~\ref{fig:sd_precoders}(c), we show the beampatterns corresponding to the first two stages as seen by the BS without any quantizer, i.e., $\left|\vc\rH\vx\right|^2$ and by the BS with 1-bit $\Sigma\Delta$ ADC, i.e., $\left|\vc\rH\vy\right|^2$, respectively, where $\vx = \sqrt{P}\alpha\va_{\rm BS}(\theta) \va_{\rm UE}\rH (\phi)\vp + \vz$ as in~\eqref{eq:mu_simo_model}, $\vy = \vx + \vq$ as in~\eqref{sd_final}, and we vary $\phi$ and $\vp$ as before. Here, we use $\theta = 30^\circ$, $\vc = \frac{1}{\sqrt{N_{\rm r}}}\va_{\rm BS}(30^\circ)$, $|\alpha|=1$, and SNR of 10 dB. We see that the received beampattern in Fig.~\ref{fig:sd_precoders}(c) is significantly distorted as the 1-bit $\Sigma\Delta$ ADC is not steered to the $\theta$, i.e., $\psi \neq \theta$, and more importantly, because the quantization voltage level $b$ is set to an arbitrary level. While we can steer the 1-bit $\Sigma\Delta$ ADC based on the AoA estimate from \texttt{Step 1} as $\psi = \hat{\theta}$, we emphasize that a procedure to select an appropriate voltage level $b$ is crucial. Before describing the procedure to select $b$, we next present the recursive bisection procedure to estimate the AoDs.

\subsubsection{Recursive bisection procedure} \label{sec:su_mimo_aod_estm}

We estimate the AoDs of the $l$th path by maximizing $E_l({\vp})$ in \eqref{eq:precoder_energy} by recursively bisecting the spatial sector and selecting a subsector that yields the highest received energy. The BS informs via a 1-bit error-free feedback link the selected subsector to the UE, which further bisects the selected subsector to transmit pilots for the next stage. The procedure is continued for $N_{\rm s}$ stages, where at the last stage, we select one of the angular grid points in~$\mathcal{D}$ as the estimated AoD of the $l$th path. We repeat this procedure for all the $L$ paths. 

To estimate the AoD of the $l$th path, we proceed as follows. Recall from \eqref{eq:precoder_condition} that $\vp_{s,i}$ has a unit response in the sector formed by departure angles $\tilde{\phi}_i \in \mathcal{I}_{s,i}$. In the first stage, the UE transmits pilots using precoders $\vp_{1,1}$ and $\vp_{1,2}$.  The BS then computes the energy of the received symbols using~\eqref{eq:precoder_energy} and selects the subsector (or the precoder) that yeilds $\max\, \{E_{l}(\vp_{1,1}),E_{l}(\vp_{1,2})\}$. If $E_{l}(\vp_{1,1}) > E_{l}(\vp_{1,2})$, the BS sends a feedback of 0 to the UE through an error-free 1-bit feedback link indicating that the sector $\tilde{\phi}_i \in \mathcal{I}_{1,1}$ is selected. Similarly, it sends a 1 indicating that the sector $\tilde{\phi}_i \in \mathcal{I}_{1,2}$ is selected. In the second stage, the UE then bisects the selected subsector and transmits pilots using the  precoders $\{\vp_{2,1}, \vp_{2,2}\}$ (respectively, $\{\vp_{2,3}, \vp_{2,4}\}$) if the received feedback from the BS is 0 (respectively, 1). More generally, at {\it Stage s}, suppose the UE transmits pilots using precoders $\{\vp_{s,m},\vp_{s,m+1}\}$ and $E_{l}(\vp_{s,m}) > E_{l}(\vp_{s,m+1})$. The BS transmits a 1 to the UE, which then selects the precoders  $\{\vp_{s+1,2m-1},\vp_{s+1,2m}\}$ corresponding to a narrower subsector for the next stage. We continue this procedure for $N_{\rm s}$ stages, where the partition selected in the final stage corresponds to the index of the estimated AoD. The same procedure is repeated for all the $L$ paths.

\subsection{Selection of clipping and quantization voltage levels}  \label{sec:su_mimo_voltage}

In Fig.~\ref{fig:sd_precoders}(c), we have seen that the choice of $b$ and $c$ play a crucial role in determining the channel estimation and beamforming performance of MIMO systems with 1-bit spatial~$\Sigma\Delta$ ADCs. As discussed in Section~\ref{sec:sigma_delta_review}, we  choose the clipping level $c$ based on the standard deviation of input that allows us to place a bound on the clipping error using the Chebyshev inequality ${\rm Pr}\left(\Vert \cL_c[\vx(t)] - \vx(t) \Vert_2 > \delta \right)  \leq \epsilon $ for constants $\delta,\epsilon >0$. Therefore, the choice of $c$ depends on the statistics of the unquantized signal $\vx(t)$ received at the BS, leading to different choices of voltage levels for \texttt{Step 1} and \texttt{Step 2} of the proposed channel estimation algorithm.

\subsubsection{For estimating AoAs and path gains in \texttt{Step 1}}
\label{sumimo_voltage_step1}

From~\eqref{sumimo:aoa_estm_received_signal} and~\eqref{sd_final}, the unquantized signal received at the $i$th antenna of the BS in \texttt{Step 1} of the proposed channel estimation technique is
\begin{equation*}
[\vx_1(t)]_i =\sqrt{\frac{P}{L}} \sum_{k=1}^{L} \alpha_k e^{-\jmath 2 \pi d (i-1) {\rm sin}(\theta_k)} + [\vz_1(t)]_i, 
\end{equation*}
where $\alpha_k$ and $[\vz_1(t)]_i$ follow a complex Gaussian distribution with zero mean and unit variance. Since the complex path gains and the additive noise are mutually independent, $[\vx_1(t)]_i$ follows a complex Gaussian distribution with zero mean and variance $P+1$. In other words, $\Re([\vx_1(t)]_i) \sim \cN(0, \frac{P+1}{2})$ and $\Im([\vx_1(t)]_i) \sim \cN(0, \frac{P+1}{2})$. Let us recall that the probability that a Gaussian random variable takes values away from the mean by more than thrice the standard deviation is less than $1\%$. Hence, to ensure
\begin{align*}
&{\rm Pr}( \vert \Re([\vx_1(t)]_i) \vert > c ) \\ 
&\quad = {\rm Pr}( \vert \Re([\vx_1(t)]_i) - \cL_c[\Re([\vx_1(t)]_i] \vert > 0 ) \leq 0.01, 
\end{align*}
in \texttt{Step 1}, we choose the clipping voltage level 
\begin{equation}
  c = 3\sqrt{\frac{P+1}{2}}.  
\end{equation}
The clipping voltage level for the imaginary part is computed similarly. The quantization voltage level $b$ is then selected using the overload condition~\eqref{eq:overload_condition}.

\subsubsection{For estimating AoDs in \texttt{Step 2}} In the second step, from~\eqref{eq:su_mimo_doa_est_idea} and~\eqref{sd_final}, 
the unquantized signal received at the $i$th antenna of the BS related to the uplink pilot transmission with the precoder $\vp$ is
\begin{equation*}
[\vx_2(t)]_i =\sqrt{\frac{P}{L}} \sum_{k=1}^{L} \alpha_k \rho_k(\vp) e^{-\jmath2\pi d (i-1) {\rm sin}(\theta_k)}  + [\vz_2(t)]_i,
\end{equation*}
where $[\vx_2(t)]_i$ follows a complex Gaussian distribution with zero mean and variance 
$\frac{P}{L}\sum_{k=1}^{L}\vert \rho_k(\vp)\vert^2  + 1$. Since $\rho_k(\vp)$ is bounded from above by $\sqrt{N_{\rm t}}$, the worst-case variance of $[\vx_2(t)]_i$ is $PN_{\rm t} + 1$. Therefore, we choose the clipping level as
 \begin{equation} \label{voltage_level_sumimo_phase2}
	c = 3\sqrt{\frac{PN_{\rm t} + 1}{2}},
\end{equation} 
in \texttt{Step 2}, to ensure that the worst-case clipping probability of $[\vx_2(t)]_i$, $i=1,\ldots,N_{\rm r}$ is less than 1$\%$. 

Since the clipping voltage level that we select in \texttt{Step 1} is different from \texttt{Step 2}, a joint estimator (e.g., sparse recovery based method as in~\cite{alkhateeb2014channel}) to jointly estimate the channel parameters is not straightforward when dealing with observations from 1-bit spatial $\Sigma\Delta$ ADCs.

\subsection{Computations and number of pilot transmissions}
\label{sec:computational_complexity}

In \texttt{Step 1}, we transmit $T_1$ pilots. In \texttt{Step 2}, we transmit each beam $T_2$ times with 2 beams in each stage. Since there are $N_{\rm s} = \log_2 D$ stages, and $L$ paths, the total pilot transmission overhead in the second stage is $2LT_2N_{\rm s}$. 

For a search grid of size $A$, computing the Bartlett beamforming spectrum in~\eqref{su_mimo_aoa_spectrum} costs about order $AN_{\rm r}T_1$ flops. The least squares estimator to compute the path gains incurs about order $L^2 N_{\rm r}T_1$ flops. In \texttt{Step 2}, for each beam and path, we compute the received power, which costs about order $LT_2 N_{\rm s} N_{\rm r}$ flops. In contrast, a scheme that exhaustively searches over all possible AoA and AoD combinations with $DT_2$ pilot transmissions incurs a computational complexity of about order $A DN_{\rm r}T_2$ flops, which is higher than $N_{\rm r}(AT_1 + LT_2 N_{\rm s})$ flops incurred by the proposed method as typically $D \gg L N_{\rm s}$.  The computational complexity of the proposed channel estimation algorithm scales linearly with the number of receive antennas and log-linearly with the search grid size, thus making it well-suited for massive MIMO systems. 

\section{MU-MIMO channel estimation} \label{sec:ch_estm_mu_simo}

In this section, we specialize the proposed channel estimation algorithm to the MU-MIMO setting with the angular channel model described in Section~\ref{subsec:mu_simo}. We estimate the channel by estimating the AoAs and the path gains using~\texttt{Step~1} of the algorithm developed in the previous section. Since we assume that the UEs have a single antenna in the MU-MIMO setup, there is no AoD estimation step.

Let $\mS(t) \in \mbC^{K \times K}$ denote the orthogonal pilot matrix transmitted at time instance $t$ from the $K$ UEs such that $\mS(t) \mS\rH (t) = \mS\rH(t) \mS(t) = \mI$. From~\eqref{eq:mu_simo_model} and~\eqref{sd_final}, the received signal at the BS with 1-bit spatial $\Sigma\Delta$ ADC is
\begin{equation*}   \label{eq:mumimo_rec_signal_basic}
\mY(t) = \sqrt{P} \mH\mS(t) + \mN(t), \, t=1,2,\cdots, T,
\end{equation*}
where $T$ is the total number of channel uses, $\mN(t)$ is the sum of additive white Gaussian noise and quantization noise, as defined before. Using the known pilots, we preprocess the received signal to separate the signal components from the UEs to obtain
$
\mY \mS\rH(t) = \sqrt{P} \mH + \mN \mS\rH(t),
$
where the $k$th column of $\mY \mS\rH(t)$, denoted by ${\vy}_{k,t}  \in \mbC^{N_{\rm r}}$, is related to the SIMO channel between the $k$th UE and the BS at time instance $t$, and is given by  [cf.~\eqref{eq:mumimo_channel}]
\begin{align*}
{\vy}_{k,t} &=  \sqrt{\frac{P}{L_k}}\mA_{\rm BS}(\boldsymbol{\theta}_k)\boldsymbol{\alpha}_k + {\vn}_{k,t} \nonumber\\
&=\sqrt{\frac{P}{L_k}}\mA_{\rm BS}(\boldsymbol{\theta}_k){\rm diag}(\boldsymbol{\alpha}_k){\bf 1} + {\vn}_{k,t}. \label{eq:mu_mimo_ch_estm}
\end{align*}
Here, ${\vn}_{k,t}$ denotes the $k$th column of $\mN \mS\rH(t)$. Suppose $\mY_{k} = [{\vy}_{k,1},\cdots, {\vy}_{k,T}]$ and $\mN_{k} = [{\vn}_{k,1},\cdots, {\vn}_{k,T}]$. Then we have
$
\mY_{k} =  
\sqrt{\frac{P}{L_k}}\mA_{\rm BS}(\boldsymbol{\theta}_k){\rm diag}(\boldsymbol{\alpha}_k){\mE} + {\mN}_{k},$
which readily resembles the signal model in \eqref{sumimo:aoa_estm_received_signal}.  Therefore, we can use~\texttt{Step~1} of the SU-MIMO channel estimation algorithm developed in the previous section to estimate the channel parameters $(\boldsymbol{\theta}_k,\boldsymbol{\alpha}_k)$ for each user. Also, we choose the same voltage level $c = 3\sqrt{\frac{P+1}{2}}$ as the one used in \texttt{Step 1} of SU-MIMO channel estimation.

\section{Numerical simulations} \label{sec:simulation}

In this section, we present results from a number of numerical simulations to demonstrate the efficacy of the developed quantization noise model, voltage level selection, and channel estimation algorithms in mmWave MIMO systems with 1-bit spatial $\Sigma\Delta$ ADCs. We compare different algorithms in terms of normalized mean square error (NMSE) and angle estimation error. We define NMSE of a path gain estimate $\hat{\boldsymbol{\alpha}}$ and a channel estimate $\hat{\mH}$ as
\[
	{\rm NMSE}(\hat{\boldsymbol{\alpha}}) = \frac{  \mbE \left[ \Vert \hat{\boldsymbol{\alpha}} - \boldsymbol{\alpha} \Vert_2^2 \right]}{  \mbE \left[ \Vert \boldsymbol{\alpha} \Vert_2^2  \right] }
\]
and
\[
{\rm NMSE}(\hat{\mH}) = \frac{\mbE\left[ \Vert \hat{\mH} - \mH \Vert_F^2 \right] }{\mbE \left[ \Vert \mH \Vert_F^2 \right]},
\]
respectively. Let $\cA=\{ \tilde{\theta}_1, \ldots, \tilde{\theta}_A \}$ denote the AoA search grid of size $A$ used in~\eqref{su_mimo_aoa_spectrum}. Let us denote the index set of AoAs corresponding to the angles in $\cA$ as 
\[
\mathbb{A}({\boldsymbol \theta}) = \{j \,: \, [{\boldsymbol \theta}]_l = \tilde{\theta}_j,  1 \leq l \leq L, 1 \leq j \leq A\}.
\]
Let us also denote the index set of AoDs corresponding to the angles in AoD grid $\cD = \{ \tilde{\phi}_1, \tilde{\phi}_2, \cdots, \tilde{\phi}_{D}\}$ of size $D$ as
\[
\mathbb{D}({\boldsymbol \phi}) = \{j \,: \, [{\boldsymbol \phi}]_l = \tilde{\phi}_j,  1 \leq l \leq L, 1 \leq j \leq D\}.
\]
We then define the AoA and AoD estimation errors as
\[
	E_\theta = {\rm Pr}\left( \mathbb{A}(\hat{\boldsymbol \theta}) \neq \mathbb{A}({\boldsymbol \theta})\right) \quad \text{and} \quad 	E_\phi =  {\rm Pr}\left(\mathbb{D}(\hat{\boldsymbol \phi}) \neq \mathbb{D}({\boldsymbol \phi})\right),
\]
where $\hat{\boldsymbol \theta}$ and $\hat{\boldsymbol \phi}$ are the estimated AoAs and AoDs, respectively.

\subsection{The SU-MIMO setting}\label{sec:simulation:sumimo}

We consider a SU-MIMO setup with the UE having $N_{\rm t}~=~32$ antennas, the BS having $N_{\rm r} = 128$ antennas that are spaced $d = 1/8$ wavelengths apart. For Bartlett beamforming, we use the search grid ${ \cA } = \{-90^\circ, -89^\circ, \cdots, 89^\circ, 90^\circ\}$ with $A = 181$ points. The AoD grid $\cD$ is obtained by uniformly sampling the direction cosine space in the interval $-1$ to $1$ with $D=128$ points as described in Section~\ref{sec:su_mimo_dod_codebook}. 
\begin{figure*}[t]
\centering
		\includegraphics[width=1.9\columnwidth]{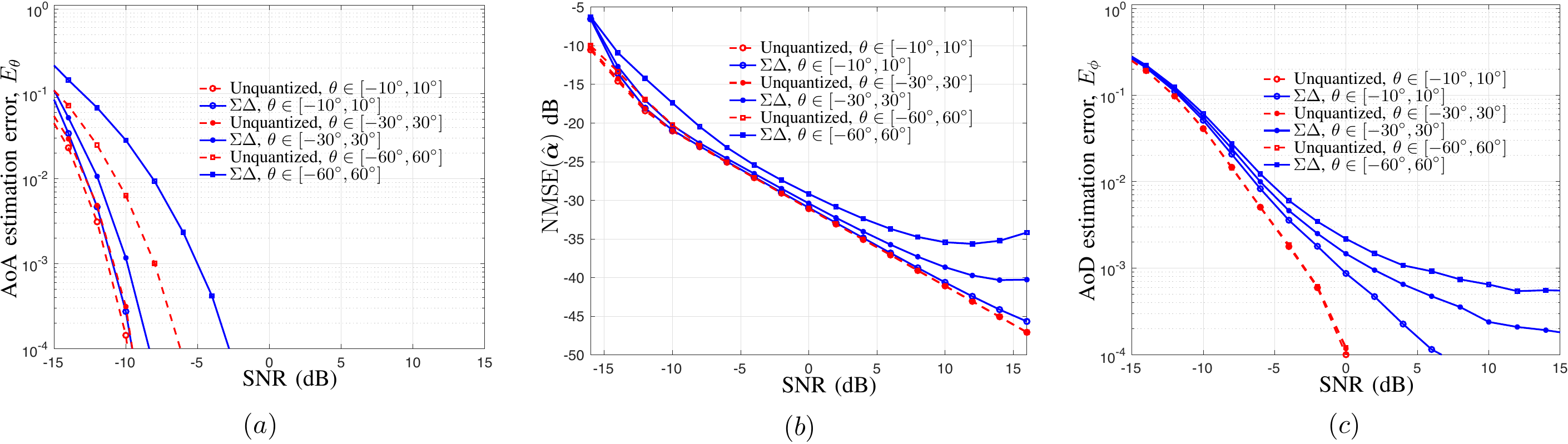}
		\caption{\small SU-MIMO performance with $L=1$. (a) AoA.
		(b) Path gain. (c) AoD.}
        \label{fig:sumimo_l1}
\end{figure*}
\begin{figure*}[t]
\centering
	\includegraphics[width=1.9\columnwidth]{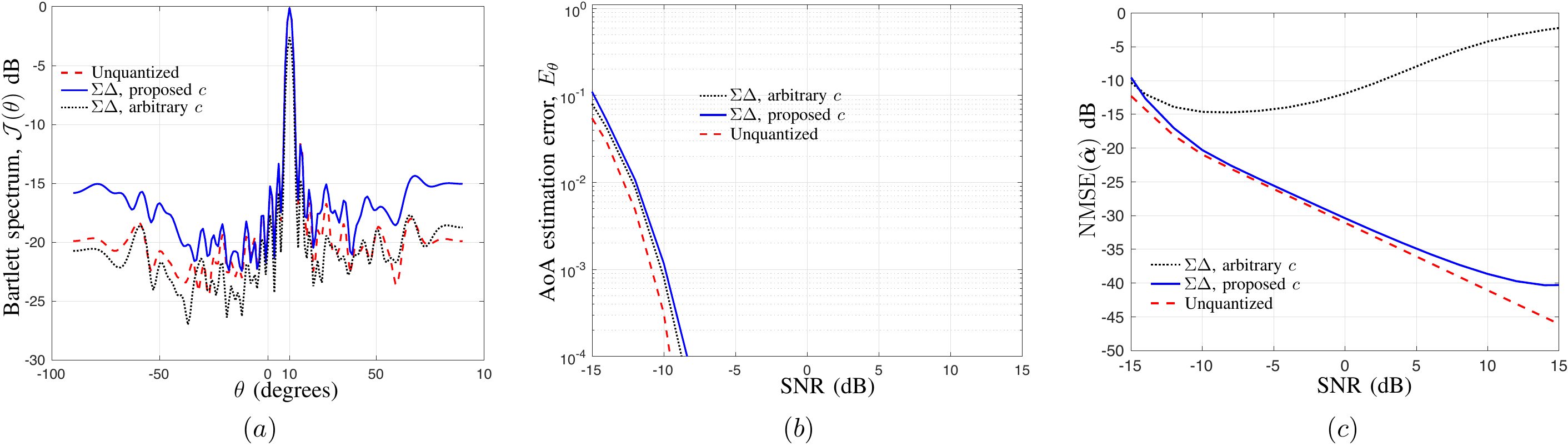}
	\caption{\small Impact of voltage level selection on \texttt{Step 1}  with $L=1$. (a) Bartlett spectrum with true AoA at $10^\circ$. (b) AoA estimation error.
		(c) Path gain estimation error.}
	\label{fig:sumimo:aoa:imperfect}	
\end{figure*}
\begin{figure*}[t]
\centering
		\includegraphics[width=2\columnwidth]{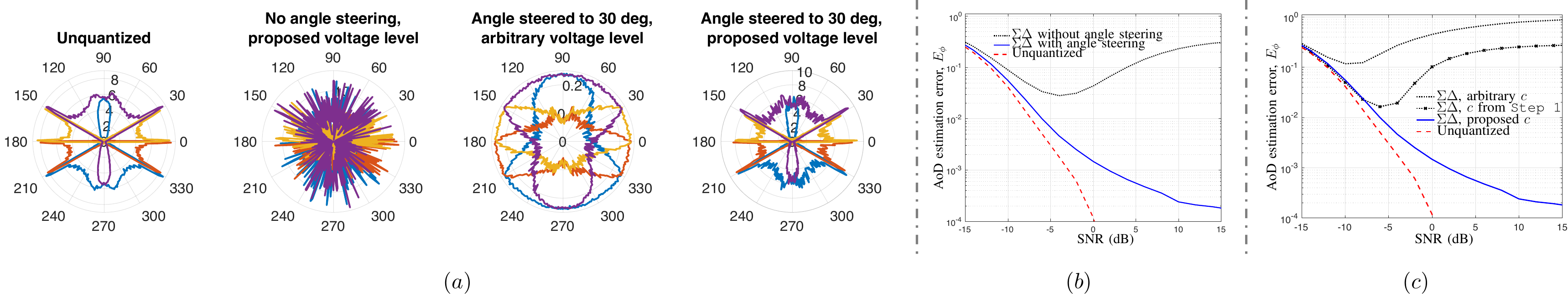}
		\caption{\small (a) Beam patterns of the \emph{Stage 2} of the proposed codebook as seen at the receiver for different voltage levels with and without angle steering.  (b) Impact of angle steering on AoD estimation. (c) Impact of voltage level on AoD estimation.}
		\label{fig:sumimo:aod}
\end{figure*}
\subsubsection{Angle steering and voltage level selection}

We first begin with a discussion on the estimation performance of the channel parameters $(\boldsymbol{\theta},\boldsymbol{\alpha})$ in \texttt{Step 1} and $\boldsymbol{\phi}$ in \texttt{Step~2} of the proposed channel estimator for a single path channel with $L=1$. This allows us to discuss the impact of voltage level selection and angle steering on the estimators. The AoA is drawn uniformly at random from the sector $[-x^\circ, x^\circ]$ for  $x=10,30,60$. The AoD is drawn randomly from the sector $[-75^\circ,75^\circ]$. The path gain is assumed to be unit modulus. We use $T_1 = 10$ and $T_2 = 1$. Unless otherwise mentioned, we use $10^6$ independent channel realizations to compute NMSE and angle errors. In this subsection, we compare the estimation performance of the proposed technique with 1-bit spatial $\Sigma\Delta$ ADC, referred to as ``$\Sigma\Delta$", with an equivalent channel estimation method applied on unquantized data (wherein the BS is assumed to have a very high-resolution ADC). We label it as ``Unquantized" in the plots. Since estimates from ``Unquantized" will be better than the same scheme applied on quantized data for a comparable number of snapshots, we use this as a baseline to illustrate the loss due to 1-bit $\Sigma\Delta$ ADC.

We illustrate the channel estimation performance for different SNRs in terms of $E_{\theta}$ in Fig.~\ref{fig:sumimo_l1}(a), ${\rm NMSE}(\hat{\boldsymbol{\alpha}})$ in Fig.~\ref{fig:sumimo_l1}(b) and $E_{\phi}$ in Fig.~\ref{fig:sumimo_l1}(c). Here, ``Unquantized, $\theta \in [-10^\circ,10^\circ]$" and ``$\Sigma\Delta, \theta \in [-10^\circ,10^\circ]$" indicate that the AoA is drawn uniformly at random from the sector $[-10^\circ,10^\circ]$ in each channel realization. We can see that ``$\Sigma\Delta$" performs similar to that of ``Unquantized" for the path with AoA arriving close to the array broadside, whereas the gap between ``Unquantized" and ``$\Sigma\Delta$" increases for the path arriving with angles away from the array broadside. This is mainly due to the quantization noise shaping at higher spatial frequencies. 
Performance degradation at higher SNRs is inevitable due to the larger choice of $b$ at high SNRs, which results in higher quantization noise [cf.~\eqref{eq:eq_noise_covariance}].

Next, in Fig.~\ref{fig:sumimo:aoa:imperfect}(a),  Fig.~\ref{fig:sumimo:aoa:imperfect}(b), and Fig.~\ref{fig:sumimo:aoa:imperfect}(c), respectively, we illustrate the impact of  clipping voltage level selection on $\cJ(\theta)$, $E_\theta$, and $E_\phi$ by comparing the proposed clipping voltage level from Section~\ref{sumimo_voltage_step1} with a fixed, SNR independent arbitrary  clipping voltage level $c=1$. While we observe that the Bartlett beampatterns as well as the AoA estimation errors are not  sensitive to the choice of clipping voltage levels, we, however, can observe from Fig.~\ref{fig:sumimo:aoa:imperfect}(c) that 
not selecting a correct clipping voltage level leads to severe performance degradation in path gain estimation. In other words,  Fig.~\ref{fig:sumimo:aoa:imperfect}(c) demonstrates that a blind application of least squares without an appropriate selection of the clipping voltage level does not result in satisfactory performance.

Recall in Fig.~\ref{fig:sd_precoders}(c), we discussed the impact of clipping voltage level selection and angle steering on the beampatterns in \texttt{Step 2} as observed at the 1-bit spatial $\Sigma\Delta$ ADC. We now extend that discussion in Fig.~\ref{fig:sumimo:aod}(a), where we focus on the second stage of the proposed codebook as observed at the receiver of an unquantized system and compare it to the beampattern as seen at the receiver with a 1-bit spatial $\Sigma\Delta$ ADC. We consider a scenario where the path arrives at the BS with an AoA of $\theta = 30^0$ and an SNR of $10$~dB. We observe that the beampatterns are severely distorted whenever the steering angle, voltage level, or both are incorrectly selected. Nevertheless, with a careful selection of voltage levels and steering angle, the beampatterns at the receiver with a 1-bit spatial $\Sigma\Delta$ ADC are comparable to that of the unquantized system.

In Fig.~\ref{fig:sumimo:aod}(b) and Fig.~\ref{fig:sumimo:aod}(c), we demonstrate, respectively, the impact of angle steering and clipping voltage level selection on AoD estimation, where the AoA is drawn uniformly at random from the sector $[-30^\circ,30^\circ]$ in each channel realization to compute $E_\phi$. We can observe that failing to exploit angle steering leads to serious loss in performance. While clipping voltage selection was not crucial for AoA estimation, we observe that fixing the voltage level to an arbitrary value, such as $c = 1$, or varying it in an inappropriate manner, e.g., using the clipping voltage level designed for \texttt{Step 1} in \texttt{Step 2} (indicated as ``$\Sigma\Delta$, $c$ from \texttt{Step1}"), do not provide satisfactory AoD estimation performance. In other words, a direct application of existing hierarchical codebook-based channel estimation methods, e.g.,~\cite{alkhateeb2014channel} without a careful selection of clipping voltage levels and phase shifts in the feedback loop will not provide reasonable performance for MIMO systems with 1-bit spatial $\Sigma\Delta$ ADCs. 

\subsubsection{Multipath channel}
We now consider a multipath SU-MIMO channel with $L = \{2,3\}$, which are typical at mmWave frequencies. We assume that the complex path gains $\alpha_i$ follow a truncated Gaussian distribution with $\vert \Re(\alpha_i) \vert , \vert \Im(\alpha_i) \vert \geq \tau$ for $i=1,\ldots,L$. To ensure that all the $L$ paths are sufficiently strong, we choose $\tau = 0.5$. The AoDs are drawn uniformly at random from the sector $[-75^\circ,75^\circ]$ with a minimum spacing of $0.1$ in the direction cosine space and the AoAs are drawn uniformly at random from the sector $[-x^\circ, x^\circ]$ with a minimum spacing of $20^\circ$, where $x^\circ = \{45^\circ,60^\circ\}$. We choose $T_1=10$ and $T_2 =1$. Recall that the proposed method with 1-bit spatial $\Sigma\Delta$ ADC, referred to as ``$\Sigma\Delta$", requires reception of $T_1$ pilots in \texttt{Step 1} and $2LT_2N_s$ pilots in \texttt{Step 2}, where the reception at each step is with a different clipping voltage level. Hence, designing a scheme that utilizes all the available $T_1 + 2LT_2N_s$ measurements for channel estimation with 1-bit spatial $\Sigma\Delta$ ADCs is not straightforward. 

We compare the performance of the proposed method with amplitude retrieval based one-bit SU-MIMO channel estimation algorithm~\cite{qian2019amplitude}, referred to as ``AR" in the plots. In ``AR", after the amplitudes are recovered from one-bit measurements, standard channel estimation algorithm can be used. Since ``AR" does not involve any clipping voltage selection, here, we use all the $T_1 + 2LT_2N_s$ measurements to perform amplitude recovery as in~\cite{qian2019amplitude}, AoA and path gain estimation using \texttt{Step 1} and AoD estimation using \texttt{Step 2} as described in Section~\ref{sec:su_mimo}. In addition to ``Unquantized", which serves as a benchmark for ``$\Sigma\Delta$" as it uses $T_1$ pilots in \texttt{Step 1} and $2LT_2N_s$ pilots in \texttt{Step 2}, we also report the performance of an unquantized system, labelled as ``Unquantized full data", that uses all the $T_1 + 2LT_2N_s$ pilots in both \texttt{Step 1} and \texttt{Step 2}. Thus, the total number of pilot transmissions in all the methods that we compare with are the same.
We have observed the runtime of the ``AR" algorithm is significantly higher than the proposed method. Due to this reason, we use $500$ independent Monte-Carlo~(MC) experiments to compute the NMSE of ``AR", whereas $40000$ MC experiments are used to obtain plots for ``Unquantized," ``Unquantized full data," and ``$\Sigma\Delta$" methods. NMSE of ``AR" is, in general, much higher than that of the other methods, which suggests that a fewer number of MC runs is sufficient to obtain curves of comparable precision.

\begin{figure}[t]
\centering
		\includegraphics[width=\columnwidth]{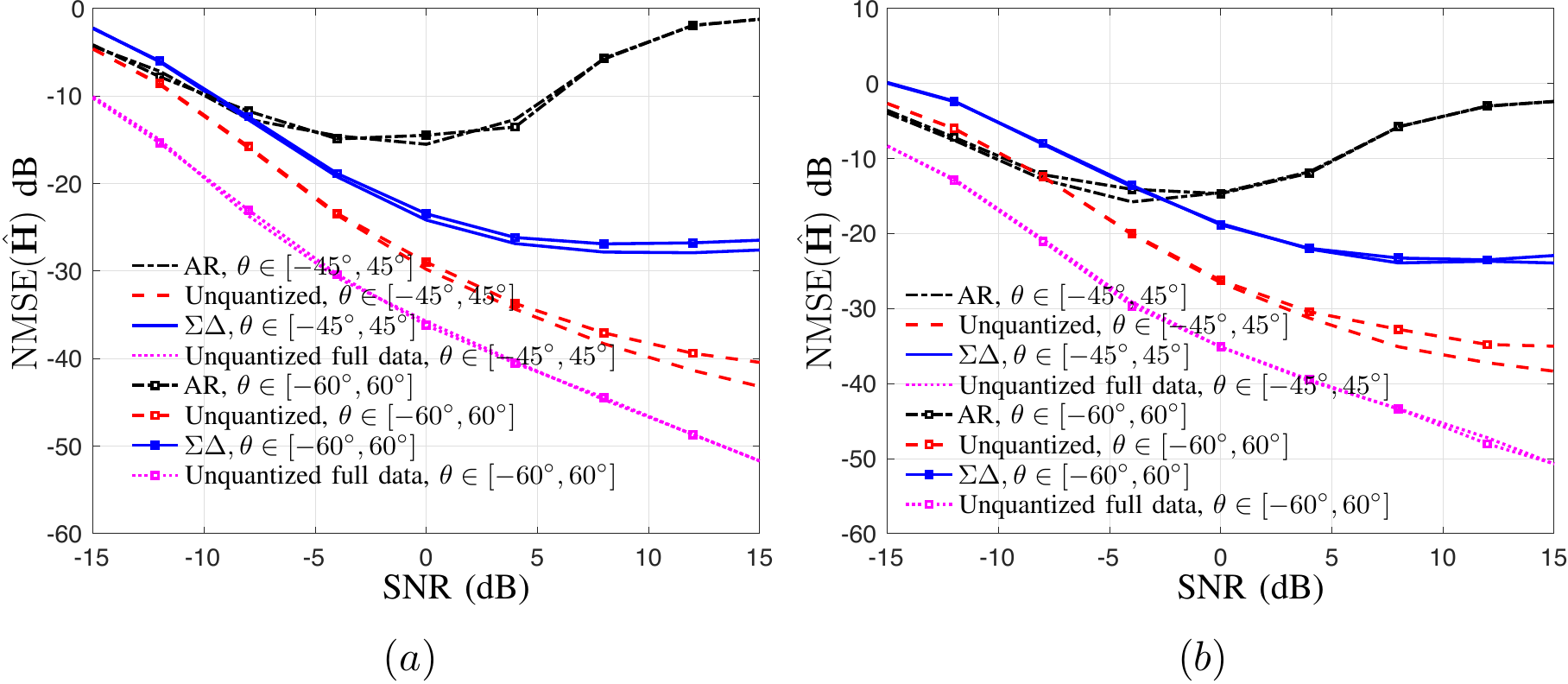}
		\caption{\small SU-MIMO channel with (a) $L=2$
		and (b) $L=3$.}
        \label{fig:sumimo}
        \vspace*{-3mm}
\end{figure}

In Fig.~\ref{fig:sumimo}, we show the channel estimation NMSE, where $\theta \in [-45^\circ,45^\circ]$  or $\theta \in [-60^\circ,60^\circ]$ indicate the sector from which the AoAs are drawn in each channel realization. We can observe that the performance of ``$\Sigma\Delta$" is comparable to that of ``Unquantized" and is better than that of ``AR" except at very low SNRs. At extremely low SNRs, we observe that the performance of ``AR" is slightly better than that of ``Unquantized" and ``$\Sigma\Delta$" as the effective number of snapshots available to estimate the channel parameters is larger for ``AR". At high SNRs, on the other hand, higher quantization noise in 1-bit systems leads to a deteriorated performance of ``AR". Similarly, at high SNRs and for angles away from broadside, there is an inevitable gap between the NMSE of ``$\Sigma\Delta$" and ``Unquantized" due to the  increase in the quantization noise. Furthermore, the NMSE of channel estimation for $L=3$ is slightly larger than that of $L=2$ due to the larger number of parameters to be estimated in the latter case. As expected, the benchmark scheme ``Unquantized full data" has the lowest NMSE due to the absence of quantization and efficient use of all available snapshots to carry out channel estimation. In essence, the proposed method achieves performance comparable to that of ``Unquantized" and significantly better than that of ``AR" for most of the SNRs, making it an attractive choice for massive MIMO systems.

 \begin{figure*}[t]
    \centering
    \includegraphics[width=2\columnwidth]{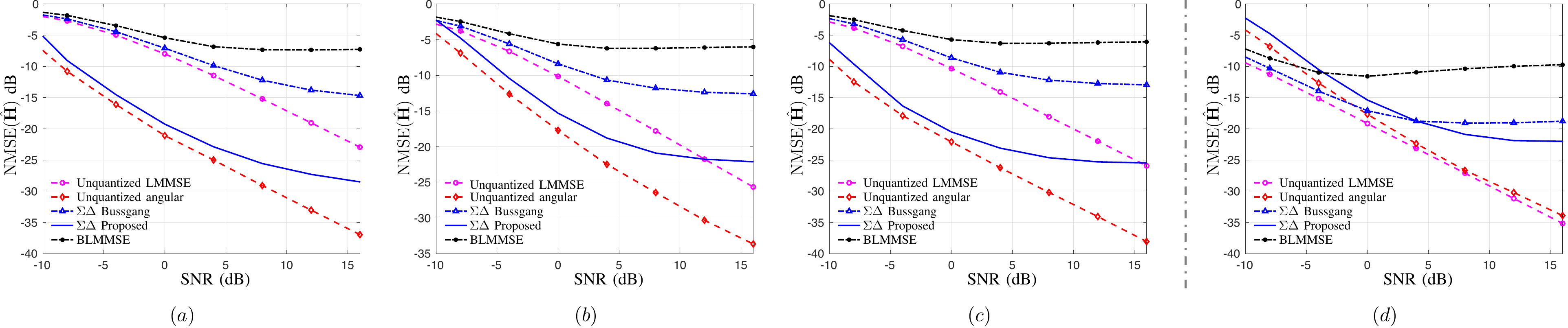}
    \caption{\small The MU-MIMO setting. (a) LoS~(i.e., $L_k =1$) with $K=8$ and $N_{\rm r} = 128$. (b) Multipath (i.e., $L_k =3$) with $K=8$ and $N_{\rm r} = 128$. (b) Multipath (i.e., $L_k =3$) with $K=8$ and $N_{\rm r} = 256$. (c) Multipath (i.e., $L_k =3$) with $K=8$ and $N_{\rm r} = 128$, where the {\it true channel correlation} is used to compute the Bussgang decomposition.}
    \label{fig:mumimo}
\end{figure*}

\subsection{The MU-MIMO setting} \label{sec:simulation:mumimo}
For the MU-MIMO setting, we consider $K=8$ users and a BS having $N_{\rm r} \in \{128,256\}$ antennas that are $d = 1/8$ wavelengths apart. We use $L_k \in \{1,3\}$ for $k=1,\ldots,8$, $T=1$, and use $500$ independent channel realizations for computing NMSE. For each path, the AoAs are drawn uniformly at random from the sector $[-45^\circ, 45^\circ]$ with a minimum spacing of $20^\circ$. We compare the channel estimation performance of the proposed method, referred to as ``$\Sigma\Delta$ {proposed}" with the following state-of-the-art techniques in  MU-MIMO channel estimation: (A) Bussgang decomposition followed by computing LMMSE MU-MIMO channel estimation for 1-bit MIMO systems~\cite{li2017channel}, referred to as ``BLMMSE", (B) MU-MIMO channel estimation with 1-bit spatial $\Sigma\Delta$ ADC based on the Bussgang decomposition~\cite{rao2020massive}, labelled as ``$\Sigma\Delta$ Bussgang", (C) an LMMSE MU-MIMO channel estimator that uses unquantized data, referred to as ``Unquantized LMMSE", and (D) the proposed channel estimation algorithm from Section~\ref{subsec:mu_simo} applied to unquantized data, referred to as ``Unquantized angular". ``Unquantized angular"  serves as the benchmark scheme for ``$\Sigma\Delta$ {proposed}" and illustrates the loss due 1-bit spatial $\Sigma\Delta$ quantization. 
 We reemphasize that ``Unquantized LMMSE", ``BLMMSE", and ``$\Sigma\Delta$ Bussgang" require the channel correlation information to perform the Bussgang decomposition and for subsequent channel estimation, and that this amounts to knowing the angles that parameterize the angular channel model. However, for the sake of comparison, we generate an {\it approximate} channel correlation by assuming that each UE has $91$ paths, each separated by 1 degree in the range of $[-45^\circ, 45^\circ]$ with ${\boldsymbol{\alpha}_k} \sim \cC\cN(0,\mI)$~[cf.~\eqref{eq:mumimo_channel}].     
 
The NMSE performance of different methods for $(L_k, N_{\rm r})=(1,128)$,  $(L_k, N_{\rm r})=(3,128)$ and  $(L_k, N_{\rm r})=(3,256)$ are presented in Fig.~\ref{fig:mumimo}(a), Fig.~\ref{fig:mumimo}(b), and Fig.~\ref{fig:mumimo}(c), respectively. It can be observed that the proposed scheme performs better in terms of NMSE than existing methods, namely, ``BLMMSE" and ``$\Sigma\Delta$ Bussgang". At low-to-moderate SNRs, the performance of ``$\Sigma\Delta$ Proposed" is comparable to that of ``Unquantized angular". This corroborates the developed theory that 1-bit spatial $\Sigma\Delta$ ADCs have higher effective resolution, which can be leveraged for parametric estimation. At higher SNRs, the NMSE performance reduces for both ``$\Sigma\Delta$ Bussgang" and ``$\Sigma\Delta$ Proposed" due to the inevitable increase in quantization noise that is bound to occur at high SNRs. Nonetheless, even at high SNRs, the proposed method outperforms ``$\Sigma\Delta$ Bussgang" by a margin of about $8~{\rm dB}$ (respectively, $12~{\rm dB}$) for $N_{\rm r} = 128$ (respectively,  $N_{\rm r} = 256$).

As expected, the performance of the parametric channel estimation techniques (i.e., ``Unquantized angular" and ``$\Sigma\Delta$ Proposed") is better than their non-parametric counterparts (i.e., ``Unquantized LMMSE" and ``$\Sigma\Delta$  Bussgang") as the parametric techniques exploit the structure in the angular channel model. As expected, the channel estimation performance of ``$\Sigma\Delta$ Proposed" and ``Unquantized angular" is better in Fig.~\ref{fig:mumimo}(a) with $L_k=1$ as compared to the multipath scenario in Fig.~\ref{fig:mumimo}(b). Furthermore, in Fig.~\ref{fig:mumimo}(a), we can see that ``$\Sigma\Delta$ Proposed" is about $8-12 ~{\rm dB}$ better when compared to the state-of-the-art method ``$\Sigma\Delta$  Bussgang". Also, when $N_{\rm r}$ is doubled from $128$ in Fig.~\ref{fig:mumimo}(b) to $256$ in Fig.~\ref{fig:mumimo}(c), due to the improved resolution and quantization noise shaping of the larger antenna array in the latter, we see that the proposed technique significantly outperforms existing techniques.

In Fig.~\ref{fig:mumimo}(d), we compare the performance of the proposed channel estimator when the true channel correlation is used to compute the Bussgang decomposition. In this setting, though not realizable in practice, ``Unquantized LMMSE" is the optimal channel estimator. We can see that ``Unquantized angular", which is the proposed technique that works with unquantized data performs similar to that of the optimal estimator at low-to-moderate SNRs. Also, we can see that the performance of our method is comparable (in terms of channel estimation NMSE) to ``$\Sigma\Delta$  Bussgang" at moderate-to-high SNRs. 

\section{Conclusions} \label{sec:conclusions}

In this paper, we have presented an algorithm for channel estimation with angular models in massive MIMO systems employing 1-bit spatial $\Sigma \Delta $ ADC. We have  developed a quantization noise model for 1-bit spatial $\Sigma\Delta$ ADCs that is useful, in general, for large array processing applications.  Although computing the complete quantization noise probability density function is difficult, the developed noise model allows us to compute its second-order statistics that can be used to prewhiten data when solving parametric estimation problems.
When the quantization voltage levels {\color{black} and phase shifts in the feedback loop} are carefully selected, the effective resolution of 1-bit spatial $\Sigma\Delta$ ADCs can be improved and hence are comparable to unquantized systems in most operating regimes of interest. We have developed a two-step channel estimation procedure to estimate the AoAs, AoDs, and path gains that characterize the MIMO channel. The proposed algorithm allows us to select the phase shifts and quantization voltage levels, which depend on the unknown channel parameters. Through numerical simulations, we have demonstrated that with the proposed channel estimation algorithm, MIMO systems with 1-bit spatial $\Sigma\Delta$ ADCs perform significantly better than MIMO systems with regular 1-bit quantization and are often on par with that of unquantized MIMO systems for low-to-moderate SNRs.
\appendix[Proof of Lemma~\ref{theo:error}]
The expressions in~\eqref{eq:sd_realimag} are obtained by adapting the derivation in~\cite{gray1989quantization} to 1-bit spatial $\Sigma\Delta$ ADCs and are derived here for self containment.

From~\eqref{eq:sd_basic_1} and~\eqref{eq:sd_basic_2} with $\psi=0$, we have the recursion 
\begin{equation} \label{eq:qn_proof_1}
	y_n(t) = x_n(t) + e_n(t) - e_{n-1}(t).
\end{equation}
Let us define the normalized quantization error as
\begin{equation} \label{eq:qn_proof_2}
	\varepsilon_n(t) = 0.5b^{-1} e_n(t) + \mu,
\end{equation}
where $\mu = 0.5 + \jmath 0.5$ and $0 \leq \Re(\varepsilon_n(t)), \Im(\varepsilon_n(t)) \leq 1$ as the quantization error $e_n(t)$ is bounded when the voltage levels are chosen as in~\eqref{eq:overload_condition}.

Using~\eqref{eq:sd_basic_1}, we can express $\varepsilon_n(t)$ and $r_n(t)$ in terms of the input and output of the quantizer as
\begin{align} 
	\Re(\varepsilon_n(t)) = \frac{1}{2b}\left( \Re (\cQ_b [r_n(t)]) - \Re( r_n(t))\right) + 0.5
	\label{eq:qn_proof_3}
\end{align}
with
\begin{equation*}
	\Re (r_n(t)) = \Re(x_n(t)) - 2b\Re(\varepsilon_{n-1}(t)) + b.
\end{equation*}
When $\Re(r_n(t)) > 0$, or equivalently, when $\Re(\varepsilon_{n-1}(t)) - 0.5b^{-1}\Re(x_n(t)) < 0.5$, we have $\cQ_b [r_n(t)] = b$. This allows us to constrain~\eqref{eq:qn_proof_3} as 
\[
0\leq \Re(\varepsilon_n(t)) = \Re(\varepsilon_{n-1}(t)) - 0.5b^{-1} \Re(x_n(t)) + 0.5 < 1.
\]
Thus, $\Re(\varepsilon_n(t)) = \langle \Re(\varepsilon_{n-1}(t) ) - 0.5b^{-1}\Re(x_n(t)) + 0.5\rangle$ as the fractional part function $\langle x \rangle = x$ for $0 \leq x < 1$ . 
Similarly, when $\Re(r_n(t)) < 0$, or equivalently, when $\Re(\varepsilon_{n-1}(t)) - 0.5b^{-1}\Re(x_n(t)) > 0.5$, we have $\cQ_b [r_n(t)] = -b$, 
\[
\Re(\varepsilon_n(t)) = \Re(\varepsilon_{n-1}(t)) - 0.5b^{-1} \Re(x_n(t)) - 0.5,
\]
and
\begin{equation}
1\leq \Re(\varepsilon_{n-1}(t)) - 0.5b^{-1} \Re(x_n(t)) + 0.5 < 2,
\label{eq:case2bound}
\end{equation}
where the lower and upper bounds are due to the fact that $0 \leq \Re(\varepsilon_{n-1}(t)) \leq 1$ and due to clipping $-b \leq \Re(x_n(t)) \leq b$.
For $x \in [1,2)$, we have $\langle x \rangle =  x - 1.$ Therefore, when $\Re(r_n(t)) < 0$, from~\eqref{eq:case2bound}, we again have
\begin{equation} \label{eq:prop1}
		\Re(\varepsilon_n(t)) = \left< \Re( \varepsilon_{n-1}(t) ) - 0.5b^{-1}\Re(x_n(t)) + 0.5   \right>.
\end{equation}
Recursively substituting for $\Re(\varepsilon_{i-1}(t))$, $\Re(\varepsilon_{i-2}(t))$, and so on in \eqref{eq:prop1}, and using the fact that $\langle  \langle x \rangle + y \rangle = \langle x + y \rangle,~\forall x,y \in \mbR$, we obtain
\begin{align*}
 	\Re(\varepsilon_n(t)) &= \left\langle 0.5(n+1) - 0.5b^{-1} \sum_{k=1}^{n} \Re(x_k(t)) \right\rangle.
\end{align*}
Using the above expression in~\eqref{eq:qn_proof_2}  yields
\begin{equation*} \label{eq:qn_proof_exp_final1}
	\Re(e_n(t)) = 2b \left\langle 0.5(n+1) - 0.5b^{-1}\sum_{k=1}^{n} \Re(x_n(t)) \right\rangle - b.
\end{equation*}
Using the fact that $\langle-x + (n-1) \rangle = - \langle x\rangle +1$, $\forall n \in \mathbb{Z}, x \in \mathbb{R}$, the above expression can be equivalently expressed  as~\eqref{eq:sd_real1}. The expression for the imaginary part in~\eqref{eq:sd_imag1} can be derived along the same lines.

\bibliographystyle{IEEEtran}
\bibliography{IEEEabrv,references}

\end{document}